\documentstyle{mn}
\onecolumn

\title[A new method of determining the inclination angle
in interacting binaries]
{A new method of determining the inclination angle in interacting binaries}
\author[Tariq Shahbaz]{
Tariq Shahbaz  \\
University of Oxford, Department of Physics, Nuclear Physics
Laboratory, Keble Road, Oxford, OX1 3RH, UK }

\begin{document}

\maketitle

\begin{abstract}
\noindent
We describe a method of determining the system parameters in
non-eclipsing interacting binaries. We find that the extent to which an
observer sees the shape of the Roche-lobe of the secondary star governs
the amount of distortion of the absorption line profiles.  The width and
degree of asymmetry of the phase-resolved absorption line profiles show a
characteristic shape, which depends primarily on the binary inclination
and gravity darkening exponent. We show that, in principle, by obtaining
high spectral and time resolution spectra of quiescent cataclysmic
variables or low mass X-ray binaries in which the mass-losing star is
visible, fitting the shape of absorption line profiles will allow one to
determine not only the mass function of the binary, but also the binary
inclination and hence the mass of the binary components.

\end{abstract}

\section{Introduction}

The determination of the binary inclination in non-eclipsing interacting
binaries has been problematic for many years. One method widely used to
determine the binary inclination in dwarf novae and the soft X-ray
transients is to measure the ellipsoidal variations of the late-type star
(Warner 1995; van Paradijs \& McClintock 1995). These variations are
caused by the companion star presenting differing aspects of its
distortion to the observer, giving rise to a double-humped modulation
whose amplitude is strongly dependent on the binary inclination. However,
one problem in measuring the ellipsoidal variations of the secondary star
is that the accretion disc can contribute a significant amount of flux at
optical wavelengths (e.g. typically about 10--50 per cent in the soft X-ray
transients). This contribution must be accounted for if the binary
inclination is to be determined using optical light curves (see Charles
1996 and references within).

Here we describe a technique which uses the information available about
the shape of the Roche-lobe of the secondary star, and its effect on the
shape of the absorption line profiles. The parameters we measure are
insensitive to the disc contribution, but dependent on the binary
inclination and gravity darkening exponent. We will first give a brief
description of the model and then describe the effect of the mass ratio,
inclination, limb and gravity darkening on the shape of the line profiles.
Finally, we will simulate data and proceed to fit it using the model.

\section{The model}

\subsection{Basic assumptions}

We model a binary system where the primary is a compact object and the
secondary a ``normal'' star. The basic assumption is that the secondary
star fills its Roche-lobe, and that its rotation is synchronised with the
orbital motion. A grid consisting of a series of quadrilaterals is set up
over the surface of the secondary star, accounting for the tidally
distorted shape in the 3-body potential of two stellar cores (i.e taking
the gravitational potential of each star to be like that of a point
mass). The equation for the Roche potential is solved numerically. Then,
for each element the radial velocity and visibility throughout the
orbital cycle are computed (see van Paradijs \& McClintock 1995). After
taking into account the intrinsic line profile, assumed to be a Gaussian
function, the predicted rotation profile at each orbital phase is
obtained by summing the line intensities of all the surface elements.
This yields the phase-resolved rotation profile of the secondary star.
The model parameters which govern the shape of the line profiles are the
binary mass ratio $q$ (=$M_{1}$/$M_{2}$ where $M_{1}$ and $M_{2}$ are the
masses of the compact object and secondary star respectively), the
inclination $i$, the mean effective temperature $T_{\rm eff}$ the mean
gravity, log~$g$, of the secondary star, and $ff$, the filling fraction
of the Roche lobe defined as the ratio of the secondary star radius to
the inner Lagrangian point distance, both being measured along the line
of centres of the two stars.

The Roche lobe filling secondary star in our model is not a normal star,
it is distorted. In the model, we assume that the absorption line arises
from a surface of constant optical depth and that that surface coincides 
with the Roche lobe equipotential. Because of gravity darkening, the two
surfaces do not exactly coincide, but for simplicity we shall take the
approximation that they do. This may introduce some uncertainty in the
specific intensity distribution over the secondary star, since
integrating over an equipotential will select different parts of the
absorption spectrum. We can estimate the validity of this assumption by
determining the scale height in the secondary star that gives rise to the
absorption line spectrum (i.e. the stellar reversing layers), and the
effect this has on the observed surface. For late-type main sequence
stars in a 10 hr binary this turns out to be 0.03 per cent of the stellar
radius (Allen 1973). Such an atmosphere will have the following effect.
If we assume that the secondary fills its critical Roche potential at
$\Omega_{crit}$ then the observed surface level will be located at
$\Omega_{crit} - \delta\Omega$. For a late-type star (K5V) in a 10 hr
binary, the observed surface lies at a potential corresponding to a star
that fills 99.97 per cent of its Roche lobe ($\delta\Omega$ is 0.03 per
cent of the stellar radius; i.e. the stellar reversing layers for a K5V
star; see Allen 1973). Using the model we calculate that there is less
than a 1 per cent change in the shape of the absorption line profiles for
a star that fills its Roche lobe compared to when it underfills it by
0.03 percent. Note however, that in general $\delta\Omega$ depends on
position on the surface, as the opacity of the stellar material is
dependent on the local temperature and gravity (Tjemkes, Zuiderwijk \&
van Paradijs 1986).

\subsection{The intensity distribution}

The distribution of the specific intensity over the stellar surface is
determined by a linear limb darkening law and the flux by von Zeipel's gravity
darkening law (von Zeipel 1924). The latter states that the radiation
flux at any point of the stellar surface is proportional to the local
acceleration of gravity. The exponent of this law depends on the
structure of the secondary star. It can be either 0.08 for a star with a
convective envelope (Lucy 1967) or 0.25 for a star with a radiative
envelope. Therefore, using Stefan's law one can obtain the local
temperature of each element of area. The local surface gravity can be
obtained by computing the derivatives of the local gravitational
potential. Given the mean effective temperature of the secondary star,
the effective temperature of each element is calculated such that the
integrated luminosity over the stellar surface is given by the observed
luminosity, which is fixed by the mean effective temperature. Similarly,
the local gravity is calculated such that the integrated gravity over the
surface of the star is given by the mean gravity of the secondary
star.

In order to determine the rotation profile, we first need to determine
the temperature and gravity of each surface element and then use model
atmospheres to determine the appropriate local specific intensity. The
line intensity is then integrated over all the visible surface elements
to give the phase-resolved rotation profile. We compute model spectra of
the Ca $\sc i$ absorption line at 6439 \AA\ for a wide range of effective
temperatures and gravities. We use this line as an example since it is
present in K-type stars, is isolated and clear from telluric features and
is strongly dependent on temperature and gravity. From these spectra we
then compute the line intensity which we store in a temperature-gravity
grid. The line intensity at a given temperature and gravity on the
secondary star can then be interpolated from this grid. For the limb
darkening at each surface element we use a linear limb darkening law, the
coefficients of which are interpolated from the values given in Al-Naimiy
(1978). However, it should be noted that the limb darkening coefficient
refers to the continuum and is different for the absorption line (see
section 3.4 and Collins \& Truax 1995). Since the line value is unknown
for late-type stars, we use the continuum value.

\section{Model simulations}

In order to show the effects of different model parameters on the shape
of the absorption lines arising from the secondary star, we compute the
rotation profiles and then convolve them with a Gaussian function,
representing the resolution of a spectrograph. Ideally one would
determine the actual response function of the spectrograph by looking at
the arc calibration spectra and then use it to broaden the rotation
profiles. In most cases the response function is Gaussian. In each test
case we assume an instrumental resolution of 6 km~s$^{-1}$ (FWHM); the
intrinsic line width is assumed to be negligible (less than the velocity
dispersion of the model profile, 2 km~s$^{-1}$; which is similar to the
turbulence velocity in late-type main sequence stars; Gray 1992). Hence
the effective line shape is only due to the combined effects of Doppler
broadening and the finite instrumental resolution, since the pixel size
is comparable to the intrinsic line width. Throughout the simulations we
will assume a secondary star with $T_{\rm eff}$=4500 K and log~$g$=4.5,
corresponding to a K5V star. We will also assume the the observed radial
velocity semi-amplitude $K_{2}$ is fixed at 162 km~s$^{-1}$
($K_{2}$=$v_{2}\sin~i$ where $v_{2}$ is the true velocity of the secondary
star).

Fig. 1 shows phase-resolved trailed spectra for three values of $\beta$
(0.0, 0.08 and 0.16) using $q$=2.0 and $i$=60$^{\circ}$. These parameters
give an expected mean rotational broadening $V_{rot}\sin i$ of 78
km~s$^{-1}$. The absorption line profiles are in the rest frame of the
secondary star. As one can see, the line profiles change shape with
orbital phase, becoming strongly asymmetric near phase 0.5 (see Fig. 5).
We measure two quantities from the model profiles, the full width at half
maximum (FWHM) and the degree of asymmetry (DOA) of the line profiles
(see Robertson 1986 for a full description). Note that the choice for the
method used to estimate the degree of asymmetry is purely arbitrary, and
only serves as a means of comparison. The degree of asymmetry is a
dimensionless parameter varying between -100 and 100. A degree of
asymmetry of zero corresponds to a symmetric profile, whereas a triangle
having one vertical side would have a degree of asymmetry of $\pm$8.1
(its sign depending on which side is the vertical one). In all the models
produced in this paper the DOA is in the range -2 to 2.

There are a number of simple tests that one can use to estimate the
numerical uncertainty in the model. For a system at zero inclination
there will be no observed radial velocity of the secondary star. The
observer will see the same projected surface of the secondary star all
the time. Also there should be no rotational broadening of the absorption
lines. Using $K_{2}$=0.0 km~s$^{-1}$ and $\beta$=0.08 for a Roche lobe
filling star we find that the DOA and FWHM are independent of orbital
phase. The DOA is 0.0 and the FWHM of the line profile is within less
than 0.1 km~s$^{-1}$ of the instrumental broadening. We obtain the same
results by setting $i$=0$^{\circ}$. 
One also expects for a spherical star that the FWHM and degree of
asymmetry of the absorption lines should be constant and independent of
orbital phase and inclination. We can simulate a spherical star by
severely underfilling the lobe of the secondary star (i.e. using
$ff$=0.1). Doing this, we find for a given set of system parameters, the
FWHM of the lines are constant to within 0.5 km~s$^{-1}$ and the DOA of
the lines are within 0.1 of zero; both the FWHM and DOA of the lines are
independent of orbital phase and binary inclination.

In Fig. 2 we show the general effects of limb ($u$) and gravity darkening
for a system at $i$=50$^{\circ}$. The case where the limb darkening ($u$)
is zero and the temperature distribution across the secondary star is
constant ($\beta$=0.0), describes the effect of the physical shape of the
Roche-lobe. Using $\beta$=0.08 changes the shape of the FWHM curves,
whereas using full limb darkening changes the mean value. 
It should be noted that the radial velocity semi-amplitude
($K_{2}=v_{2}\sin~i$) depends on the true velocity of the secondary star
($v_{2}$) and the binary inclination. Since $v_{2}$ only sets the
absolute velocity scale of the system, in subsequent sections we will
assume the observable parameter $K_{2}$ to be fixed, so that we can also
compare the FWHM of the line profiles for different model parameters.
Below we describe in detail the effects of the various model parameters
on the shape of the line profiles.

\subsection{The effects of inclination}

We find that for a given gravity darkening exponent, the shape of the
FWHM curves, especially near phase 0.4, are dependent on the binary
inclination. At low inclinations the observer generally sees the same
surface elements on the star, and so the FWHM light curves will be almost
symmetric around phase 0.25. The largest value for the FWHM of the line
profiles will be for the profiles at phase 0.25, since at this phase the
observer sees the full length of the distorted star, with the largest
range of velocities. As the binary inclination increases, the shape of
the light curve changes. The general shape of the light curve up to phase
0.25 is the same, however, after phase 0.25 the profiles start to change
significantly because the observer starts to see a larger fraction of the
inner distorted face of the star. This results in the line profiles
becoming highly asymmetric because of the heavy limb and gravity
darkening associated with the surface elements on the inner face of the
star. Fig. 3 shows the FWHM and DOA curves for fixed values of
$\beta$=0.00 and $\beta$=0.08 with $i$=20$^{\circ}$, 40$^{\circ}$ and
60$^{\circ}$. As the binary inclination increases, the degree of
asymmetry of the line profiles near phase 0.45 shifts to more negative
values. In Fig. 3c we show in detail the effects of the binary inclination
on the DOA of the line profiles, for low binary inclinations
$i$=0$^{\circ}$, 2$^{\circ}$, 4$^{\circ}$, 8$^{\circ}$, 15$^{\circ}$
using $\beta$=0.08. The asymmetry near phases 0.05 and 0.45 first shifts
to more positive values as the inclination decreases. At the maximum DOA
value, at $i\sim 15^{\circ}$, the DOA of the profile at all orbital
phases then moves towards zero; at $i=0.0 ^{\circ}$ the DOA of the phase
resolved line profile are zero. From Fig. 3a and 3b it can also be seen
that the minimum of the FWHM curves at phase 0.5 shifts towards phase
0.4. This shift of the minimum of the FWHM curves can be explained in
terms of the observer seeing a larger fraction of the distorted inner
face of the secondary star. Note that the degree of asymmetry of the line
profiles near phase 0.25 is almost independent of the binary inclination.
This is simply because the observer sees the full extent of the Roche
lobe, which is almost the same for all inclination angles.

\subsection{The effects of gravity darkening}

As mentioned earlier the temperature distribution across the secondary
star is governed primarily by the gravity darkening exponent. If
$\beta$=0.0 then each surface element will emit the same flux. However,
the inclination, line of sight effects and the shape of the Roche-lobe
then govern the final flux distribution one observes. If $\beta$ is
large, then the surface elements with the least gravity, i.e. around the
inner Lagrangian point, will be heavily gravity darkened. Even the
surface elements near the spherical face are gravity darkened. These
elements will have much lower temperatures than the elements near the
pole of the star and thus will emit less flux. This implies that the
cores of the line profiles at phases 0.0 and 0.5 will appear to be filled
by a small amount, since these cool elements of area have low velocities.
Fig. 4 shows the temperature distribution along the line of centres of the
two stars, for $\beta$=0.0, 0.08 and 0.16.

The effects of gravity darkening can seen in Fig. 1. Here we show the
profiles for $\beta$=0.0, $\beta$=0.08 and $\beta$=0.16. As one can
see, the asymmetry in the profiles near phase 0.45 changes dramatically.
Fig. 5 shows the FWHM and DOA curves for a system at $i$=20$^{\circ}$
and $i$=60$^{\circ}$ with $\beta$=0.00, 0.08 and 0.16. As the gravity
darkening exponent increases, the degree of asymmetry of the line profiles
near phase 0.45 shifts to more negative values. For high
inclination systems the minimum of the FWHM curves at
phase 0.5 shifts towards phase 0.4 (a similar effect to that observed 
by increasing the binary inclination).

\subsection{The effects of the mass ratio}

Fig. 6 shows a system with $i$=60$^{\circ}$, $\beta$=0.08 and $q$=1
(stars), 5 (circles) and 10 (crosses). The main effect $q$ has is to
change the FWHM of the line profiles; 
the FWHM changes fastest with $q$ for small values of $q$.
The DOA of the profiles do not change much. $q$
governs the amount of distortion and hence the shape of the Roche-lobe.
However, the rate of change of the distortion is greatest at low values
of $q$. In Fig. 7 we measure the amount of distortion by computing the
ratio of the Cartesian coordinates $x/y$, $x/z$ and $z/y$ at the surface
of the secondary star, where $x$ is defined to lie along the line joining
the centre of mass of the two stars, $y$ to lie in the orbital plane and
$z$ perpendicular to the orbital plane, i.e along the axis of rotation.
In fact, at extreme values of $q$, the amount of distortion of the
secondary star tends towards a constant value, which explains the
apparent convergence of the FWHM curves at large $q$.

\subsection{The effects of limb darkening} 

In the model, we use values for the limb darkening coefficient which
refer to the continuum (Al-Naimiy 1978), since the actual value, which
varies in the absorption line, is unknown for late-type stars. We assume
a local intrinsic line broadening due to turbulence of 2 km~s$^{-1}$
(i.e. the velocity dispersion of the data), which corresponds to a change
in wavelength of $\sim$ 0.04\AA. For a given temperature we find that the
continuum limb darkening coefficient varies by less than 0.01 per cent
(from the wavelength dependence of the limb darkening coefficient;
Al-Naimiy 1978) in the intrinsic absorption line profile. The effect of
this change is negligible compared to the change in the limb darkening
coefficient ($\sim$ 15 per cent) as a result of the temperature
variations across the star.

The effects of limb darkening on the mean value for the rotational
broadening of the secondary star have been noted before. Welsh, Horne \&
Gomer (1995) find that the mean broadening changes by up to 14 percent,
between zero and full limb darkening. Here we explore the effects of the
actual value used for the local limb-darkening coefficient (for each
element on the surface of the star) and how its value changes the shape
of the line profiles, in particular the shape of the FWHM curves. Fig. 8
shows the effect of zero ($u$=0.0) and full ($u$=1.0) limb darkening on
the shape of the FWHM curves, for $i$=30$^{\circ}$ and $i$=50$^{\circ}$
with fixed $\beta$=0.08. The FWHM curves have been normalised to the
value at phase 0.25. Note that the whole shape of the FWHM
curves changes between full and zero limb darkening, especially near
phase 0.0 and 0.5.

\section{The effect of zonal flows and star-spots}

\subsection{Zonal flows}

Rapid rotation, strong tidal force and non-uniform heating are present in
all interacting binaries. These effects give rise to circulation currents
within the atmospheres of the mass losing stars. The results of Martin \&
Davey (1995) suggests that for temperature rise of 1000K the
predicted flows are comparatively slow (less than 1 km~s$^{-1}$). One
would expect such low velocities, since the sound speed of material
in the outer layers of the star is $\sim$7 km~s$^{-1}$ (Frank, King \&
Raine 1992). We have simulated such flows by
adding a random velocity between 0 and 1 km~s$^{-1}$ to the radial velocity
of the elements on the surface of the star. 
The observed profile is determined as before (see section 2.1). 
We find that although the shapes of the line profiles do change, the
general trends in the FWHM and DOA curves are very similar to the
simulations for a star without the zonal flows (see Fig. 9). Therefore we
conclude that the effects due to the gravity darkening and inclination
are dominant. However, it should be noted that we have assumed a random
velocity flow pattern; the exact 3-D velocity pattern is probably very
complicated, but if we use a flow pattern that is restricted around the
equator of the star, the result does not change.

\subsection{Star-spots}

In CVs tidal dissipation is likely to produce uniform rotation as well as
synchronous rotation. Atmospheric motions have been used to explain the
anomalous brightness distribution on the surface (a hot-spot near the
inner Lagrangian point) on some of the irradiated secondary stars in
cataclysmic variables (Davey \& Smith 1992). The simulations of Martin \&
Davey (1995) show hot buoyant material rising at phase 0.4 and sinking
near phase 0.9 and phase 0.0. They then go on to suggest that the
converging flow at the back of the star may gather up magnetic field
lines and so produce a magnetic star-spot (dark-spot).

In order to estimate the effects of hot- and dark-spots on the asymmetry
of the line profiles, we simulated absorption line profiles for a star
with a dark-spot centred at the back side of the star and a hot-spot
centred at the inner Lagrangian point. Martin (1988) showed that the
position of the white dwarf and bright spot on the accretion disc in
cataclysmic variables would give rise to irradiation that was not too far
from being symmetric about the inner Lagrangian point. Therefore we
describe the spot as a circle extending to a latitude R$_{\rm
spot}$$^\circ$ away from its centre (i.e. the inner Lagrangian point for
the hot-spot and the back-side of the star for the dark-spot). The
effective temperature in the dark-spot region is lower than it would have
been in the absence of the spot by 750 K. This is the typical observed
temperature difference in RS CVn systems and T Tauri stars (Rodono 1986).
We have used a hot spot which has an temperature increase of 1000 K, but
note that this is an extreme case. For dwarf novae or soft X-ray
transients in quiescence the temperature increase at the inner face of
the secondary is probably much less. We have used spot sizes of 10
degrees.

Fig. 10 shows FWHM and DOA curves using $i$=60$^{\circ}$ and $\beta$=0.08
and spot sizes of 10 degrees, for a star with a dark- and hot-spot and
with no spots. As one can see, only the dark-spot has an appreciable
effect on the shape of the line profiles near phase 0.05. However, its
effect depends on its size. It should be noted that the simulations shown
here are an extreme case; no dark-spot at the back side of the secondary
star has been observed in any CV.

\section{Simulating data} 

In this section we describe the method we use to determine the system
parameters from simulated data. We determine the binary inclination, the
gravity darkening exponent and also the fraction of light arising from
the secondary star.

In order to simulate data, we computed 21 rotation profiles using
$i$=60$^{\circ}$, $\beta$=0.08, $q$=2.0, $K_{2}$=162 km~s$^{-1}$ and
$T_{eff}$=4500 K at orbital phases between 0.0 and 0.5. We also allowed
for the effects of a velocity flow across the star (see section 4.1). We
then convolved these functions with a template star spectrum (a K5 V
star) and then added Gaussian distributed noise. The observed template
star spectrum was taken using the Utrecht Echelle Spectrograph (FWHM=6
km~s$^{-1}$). The effects of an accretion disc were simulated by veiling
the observed flux by 33 per cent.

To estimate the typical signal-to-noise we might expect to achieve for
each spectrum, we use the dwarf nova AE Aqr as an example. Observing this
object with the high resolution spectrograph attached to the Keck
telescope for 6 mins would give a signal-to-noise ratio of about 80 per
pixel (using the program $\sc hires$ with V=12 mags at 6500 \AA). In
practice, the absorption line features will be smeared by an amount $2
\pi K_{2}t \cos(2 \pi \phi)/P$, due to the orbital motion of the
secondary star during the length of each exposure ($t$) at orbital phase
$\phi$. This effect is included by convolving the phase resolved broadened
template star spectra with a rectangular function with the appropriate
velocity width. In this example the effect of smearing is about 10
km~s$^{-1}$, i.e. t $\sim 0.01P$. The final phase resolved simulated data
have a velocity dispersion of 2 km~s$^{-1}$. The spectra were then
Doppler shifted into the rest frame of the secondary star.

It should be noted that in simulating data we first applied the
instrumental broadening to the absorption line spectrum and then
convolved it with the rotation profile of the secondary star. However, in
reality the absorption line spectrum is first broadened by the rotation
of the secondary star and then by the response of the instrument. Using
Gaussian and Lorentz instrumental response functions and a highly
asymmetric rotation profile (i.e. one at phase 0.45) we performed the
convolutions in the orders described above. We find that the maximum
difference in the convolved profiles is less than 0.02 per cent. It will
clearly be hidden in the noise of the simulated data. This small
difference is due to the fact that the intrumental broadening is more
than a factor of 10 less than the rotational broadening of the secondary
star.

\subsection{Measuring $\bf V_{\bf rot}\sin i$}

Tidal synchronisation will make the angular velocity of the secondary
star constant. Since the star is distorted, the linear rotational
velocity will vary with longitude around the star and the true
$V_{rot}\sin i$ will vary with orbital phase. Many authors have attempted
to measure the changes in the phase resolved rotational broadening in
cataclysmic variables (Welsh, Horne \& Gomer 1995; Casares et al. 1996).
In particular Casares et al. (1996) determined it in AE Aqr. They compare
the $V_{rot}\sin i$ variations with the FWHM of model line profiles.
However, it should be noted that there are many systematic errors in the
procedure they use. The method to extract $V_{rot}\sin i$ is flawed, i.e.
using the Gray profile (Gray 1992) to broaden a template star spectrum
which is then compared with the data, because the secondary star is
non-spherical. In this section, we calculate the systematic effects
introduced by only fitting the $V_{rot}\sin i$ variations with the
FWHM curves. Note that if one wants to determine the binary inclination
then one has to fit the actual shape of the absorption lines using a
model that predicts the correct broadening function for a star that fills
its Roche-lobe.

First, we extract the $V_{rot}\sin i$ variations from the simulated data.
Note that the specific intensity versus temperature and gravity relations
we use in the simulated data are for the total intensity in the spectral
range 6380--6460 \AA. We broadened the template star spectrum from 90 to
100 km~s$^{-1}$ in steps of 2 km~s$^{-1}$ using the Gray rotation profile
(Gray 1992). A linear limb-darkening coefficient of 0.65 was used
(Al-Naimiy 1978). We then performed an optimal subtraction between the
phase resolved broadened template and data spectra. The optimal
subtraction routine subtracts a constant times the template spectrum from
the data, adjusting the constant to minimize the residual scatter between
the spectra. The scatter is measured by carrying out the subtraction and
then computing the $\chi^{2}$ between this and a smoothed version of
itself. The constant, $f$, represents the fraction of light arising from
the template spectrum, i.e. the secondary star. The optimal values of
$V_{rot}\sin i$ and $f$ are obtained by minimising the $\chi^{2}$. Note
that the data point at phase 0.475 (Fig. 11) has a larger velocity width
(FWHM) than its neighbours. This is because, at this phase, the line
profile is highly asymmetric (see section 3), and so the method of using
the Gray profile fails badly. The next step is to determine the
model FWHM line profile variations. We do this by determining the
rotation profile for a given set of model parameters, i.e. inclination
and gravity darkening exponent. We then convolve this with a Gaussian
function representing the instrumental resolution. The FWHMs of the phase
dependent line profiles are then measured.

The measured phase resolved values of $V_{rot}\sin i$ are then fitted
with the FWHM variations of the line profiles. We exclude the data point
at phase 0.475, since it is clearly incorrect. This allows us to
perform a grid search in the $\beta-i$ plane, allowing confidence levels
to be determined. We obtain a best fit at $i$=42$^{\circ}$ and
$\beta$=0.14. Fig. 11 shows the measured phase resolved $V_{rot}\sin i$
variations and the best model fit. In Fig 12 we show the 68 and 90 per
cent confidence levels of the fits in the $\beta-i$ plane. As one can
see, even at the 90 per cent confidence level, we do not recover the
original system parameters of the simulated data. There seems to be a
systematic shift towards lower values for the inclination and larger
values for the gravity darkening exponent. This result is not so
surprising if we consider the fact that we are only using the widths of
the lines. One can see from the model simulation in section 3, that at
certain orbital phases, i.e. 0.1 and 0.4, the widths of the line profiles
may be the same, but the actual shapes of the profiles are not. One has to
fit the actual shape of the absorption lines using a model that predicts
the correct broadening function for a star that fills its Roche-lobe.

\subsection{Fitting the line profiles}

In principle, one would like to use many absorption lines in the fitting
procedure. However, the advantages of using single line profiles can be
seen by comparing the normalised line intensity of the Ca $\sc i$ line
with the total specific intensity in the spectral range 6380--6460 \AA\
(see Fig. 13). Generally, using a wide spectral range will tend to
include absorption lines that are not so sensitive to temperature and
gravity. The net effect will be to decrease the sensitivity of the
fitting procedure. Fitting many strongly temperature and gravity
dependent single lines will make the fitting procedure more sensitive. In
this section we show that we can recover the binary inclination by only
fitting a single line.

In order to fit the phase resolved simulated spectra, we first determine
a rotation profile for a given set of model parameters, i.e. orbital
phase, inclination and gravity darkening exponent, which we then convolve
with the template star spectrum (i.e. the isolated Ca $\sc i$ line). Note
that the template star spectrum has already been broadened by the finite
resolution of the spectrograph (6 km~s$^{-1}$), so we do not have to
allow for this in the model. We then perform an optimal subtraction
between the phase resolved broadened template and data spectra. The
values of $\chi^{2}$ for each orbital phase are then summed up. This
allows us to perform a grid search in the $\beta-i$ plane, allowing
confidence levels to be determined. We obtained the best fit at
$i$=66$^{\circ}$ and $\beta$=0.06. 

We have also simulated data with a signal-to-noise ratio of 80 per
spectrum, as previously, but now with a 10 degree hot- and dark-spot
placed at the inner Lagrangian point and the back side of the secondary
star respectively, and a random velocity flow pattern (see section 4)
over the star. We find that it only changes the inclination solutions by
8 degrees; the 1-$\sigma$ uncertainty in the model fit is 12 degrees.
Therefore within the uncertainties the star-spots and zonal flows do not
have a significant effect on the inclination angle determined.

In the data simulation we have assumed the instrumental response to be a
Gaussian function; for most spectrographs this assumption is good.
However, in order to see the effects of fitting data taken with a
spectrograph that does not have a Gaussion instrumental response, we
simulated data using a skewed Gaussian function as the instrumental
response. The function is a normal Gaussian, except that we multiply the
FWHM of half the curve by ($\sinh \gamma /(1 - \exp -\gamma $). This means
that one side of the curve is the same as a Gaussian with that FWHM, and
the other side is skewed. In this case we used $\gamma$=2.0; using
$\gamma$=0.0 would give a Gaussian profile. We then went on to fit the
data, using a model with a Gaussian function as the instrumental
response, and find that it only changes the inclination solutions by at
most 3 degrees.

It should also be noted that we have used continuum limb-darkening
coefficients in the model. The absorption lines in late-type stars will
have core limb-darkening coefficients much less than the value
appropriate for the continuum; the precise value requires detailed
calculations (Collins \& Truax 1995). However, in order to estimate the
uncertainties in over-estimating the limb-darkening in the line we
simulated data using limb-darkening coefficients which were 10 percent of
the appropriate continuum value. We then proceeded to fit the data using
the model described. We find that using the model in which the
limb-darkening coefficients is the appropriate continuum value, the
1-$\sigma$ uncertainty in the inclination angle solution is 15 degrees,
the best fit is at $i$=66$^{\circ}$. Using the model in which the
limb-darkening coefficients is 10 percent of the continuum value,
the 1-$\sigma$ inclination solutions are 4 degrees higher, $\beta$ is
0.02 higher. Therefore, despite using a model in which we over-estimate
the limb-darkening coefficient in the absorption line, we can still
determine the binary inclination.

If we assume that we can fit 16 single line profiles that have similar
intensity versus temperature and gravity relations to those for the Ca $\sc i$
line, then our uncertainties are reduced by a factor 4. In Fig. 14 we
show the 68 per cent confidence level of the fits using 16 lines in the
$\beta-i$ plane. The cross marks the best fit and the star marks the true
parameters of the simulated data. For $\beta$=0.08, we find
$i$=50$^{\circ}$--68$^{\circ}$ (68 per cent confidence). In Fig. 15 we
show the phase resolved values of $\chi^{2}_{\nu}$ and the veiling factor
($f$, the fraction of light arising from the secondary star) obtained
using the best fit model. As one can see, we have demonstrated that by
fitting a single line profile, one can at least recover the binary
inclination. It should be noted that the accuracy with which we can
determine the inclination is dependent on the signal-to-noise of the
phase resolved spectra. In this example it is 80 in a single observation.
This can easily be increased by observing the binary system for two
orbital cycles and then coadding the spectra on half the orbital period,
since the spectra between phases 0.5 and 1.0 will be symmetric to those
between phase 0.0 and 0.5. This will increase the signal-to-noise of each
individual spectrum by a factor 2. Also note that simultaneously fitting
many single lines that are correlated with temperature and gravity in
different ways may allow one to also extract the gravity darkening
exponent. However, this involves a detailed examination of model line
profiles in order to select single lines which have different temperature
and gravity relations. This will be the subject of a future paper.

\section{Discussion and Conclusions}

By obtaining high spectral and time resolution spectra of cataclysmic
variables or low mass X-ray binaries, systems in which the absorption
features of the secondary star can be seen, one can perform a radial
velocity study of the secondary star and also determine the binary mass
ratio (see Marsh, Robinson \& Wood 1994 for a full description of this
kind of study). However, as pointed out in this paper, it is also
possible to extract the binary inclination by fitting the absorption
lines profiles of the secondary star.

Some of the assumptions inherent in the model which predicts the
shape of the line profiles should be noted. The main assumption in the
model is that the observed surface of constant optical depth coincides
with the Roche potential surface. However, we find that this only changes
the shape of the line profiles by less than 1 per cent. Zonal flow patterns
and/or dark- hot-spots can have an appreciable effect on the shape of the
line profiles. However, the extent of these effects depends very much on
the magnitude of the zonal flows and the size of the spots. Simulations
show that only very low velocities are predicted for zonal flows 
(Martin \& Davey 1995).

The characteristic modulations in the width of the line profiles are, in
principle, similar to the ellipsoidal modulations, which are due to the
observer seeing differing aspects of the gravitationally distorted
secondary star as it orbits a compact object. However, this method uses
the velocity information across the star as well as the projected area.
It should also be noted that unlike modelling the ellipsoidal variations
in the optical, where the accretion disc contribution must be taken into
account, this method is independent of the disc contribution (as long as
the disc contribution is not such that it totally swamps the secondary
star features). This implies all emission and continuum sources, i.e.
from a bright spot or a gas stream, will only affect the determination of
the veiling factor and not the shape of the absorption lines. However,
care must be taken in choosing the lines to use for this kind of
analysis; the lines must be clear of any weak emission features arising
from the disc or accretion flow.

The determination of the binary inclination is essential if one wants to
obtain the mass of the binary components in non-eclipsing binaries. The
method described here can be applied to bright cataclysmic variable stars
or low mass X-ray binaries, in which one can resolve the absorption lines
of the secondary star, and where the amount of X-ray heating is small.
The effect of heating implies that one cannot use temperature sensitive
absorption lines such as the Na $\sc i$ 8183-8184 \AA\ doublet in the
modelling (see Friend et al. 1990 and references within for the effects
of irradiation). Such systems are the dwarf novae and the soft X-ray
transients. This method allows one to determine all the system parameters
from a high spectral resolution (few km~s$^{-1}$) spectroscopic study.

We have described a method of determining the binary inclination in
non-eclipsing interacting binaries by fitting the shape of the absorption
lines arising from the secondary star. We find that the amount of
distortion of the absorption line profiles is primarily due to the extent
to which an observer sees the shape of the Roche-lobe of the secondary
star. We show that, in principle, by obtaining high spectral and time
resolution spectra of quiescent dwarf novae or the soft X-ray transients,
where the disc is low, fitting the shape of absorption line profiles will
allow one to determine the binary inclination. Our simulations show that
previous efforts to determine the inclination are flawed, and give
systematically lower values for the inclination.

\section*{Acknowledgements}

\noindent
I would like to thank Bill Welsh and Phil Charles for valuable
discussions, and Barry Smalley for computing the model atmospheres. I
also thank the referee R.C. Smith for his careful reading and criticism
of the paper that has undoubtedly enhanced its content. This work was
carried out on the Oxford Starlink node using the $\sc ark$ software
package to plot the figures.

\section*{Figure captions}

\noindent
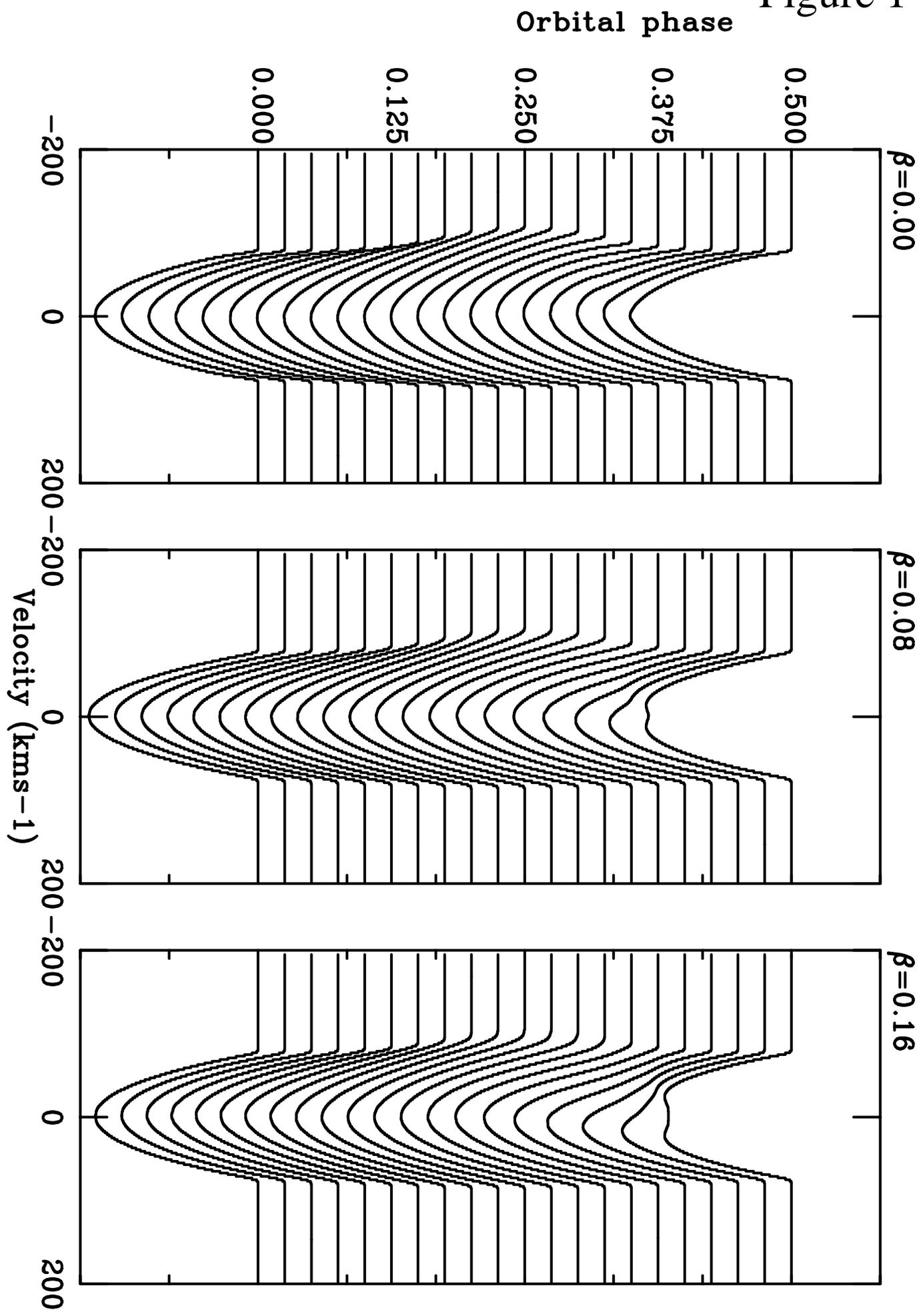
{\bf Figure 1:} From left to right, the panels show the trailed
absorption line profiles using $\beta$=0.0, 0.08 and 0.16. For all the
models $q$=2.0, $i$=60$^{\circ}$ and $K_{2}$=162 km~s$^{-1}$. As one can
see the width and the degree of asymmetry of the line profiles are
dependent on orbital phase and the value for $\beta$.  \\

\noindent
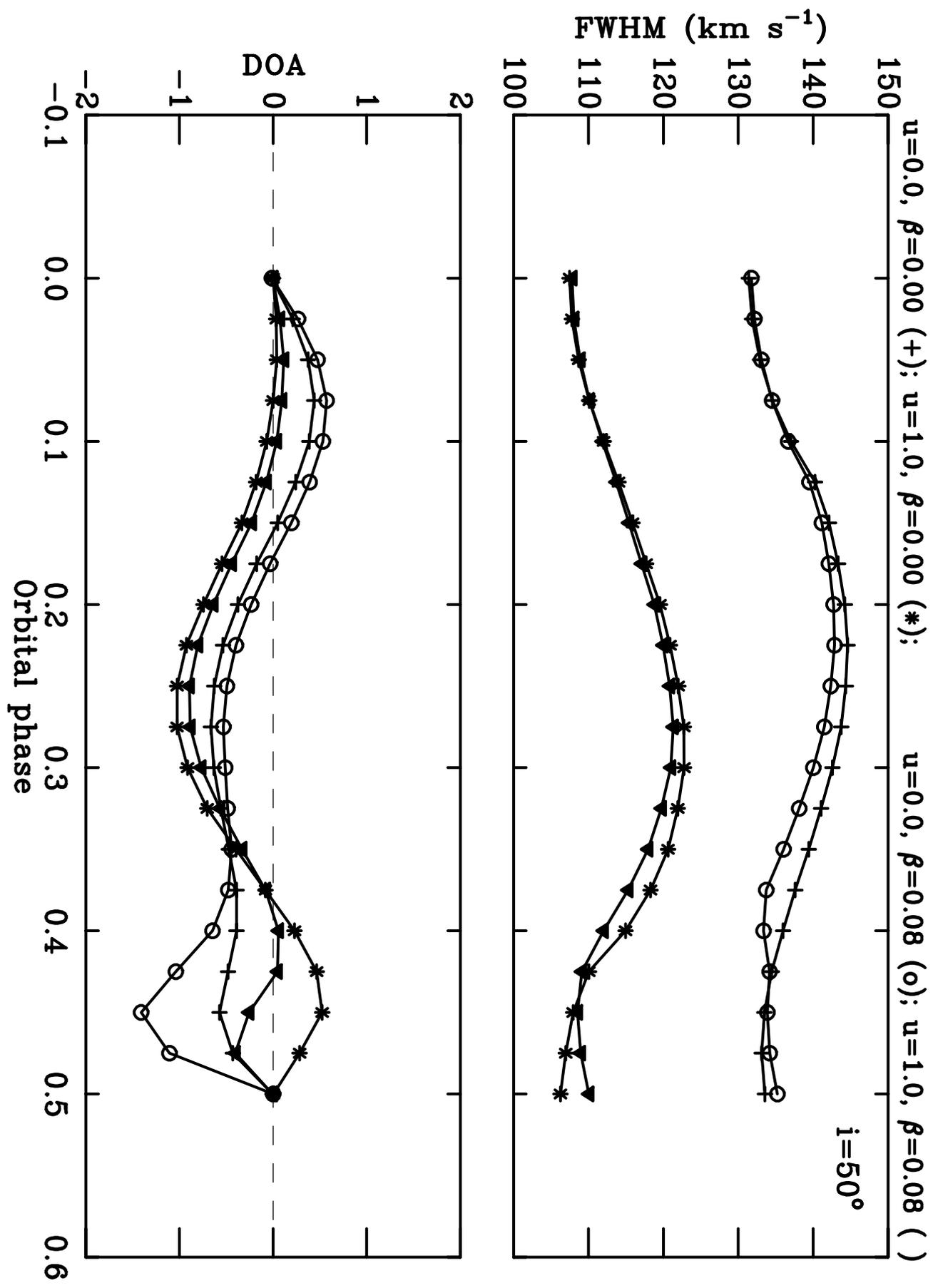
{\bf Figure 2:} The general effects of limb and gravity darkening on the
FWHM and DOA curves for a system at $i$=50$^{\circ}$. The case for
$u$=0.0, $\beta$=0.00 (plus signs); $u$=1.0, $\beta$=0.00 (stars);
$u$=0.0, $\beta$=0.08 (circles); $u$=0.0, $\beta$=0.08 (triangles) are
shown. The gravity darkening changes the shape of FWHM curves whereas the
limb darkening changes the mean value. The horizonal dashed line
indicates where the degree of asymmetry of the line is zero.  \\

\noindent
{\bf Figure 3:} The effects of inclination. 3(a) The FWHM and DOA curves
for $\beta$=0.00. Models for $i$=20$^{\circ}$ (stars), 40$^{\circ}$
(circles) and 60$^{\circ}$ (crosses) are shown. The horizonal dashed line
indicates where the degree of asymmetry of the line is zero. 3(b) Same as
above but using $\beta$=0.08. 3 (c) The DOA curves for low binary
inclinations, 0$^{\circ}$, 2$^{\circ}$, 4$^{\circ}$, 8$^{\circ}$,
15$^{\circ}$, 30$^{\circ}$ and 60$^{\circ}$, using $\beta$=0.08. As the
binary inclination decreases the degree of asymmetry of the line profiles
near phases 0.05 and 0.45 shifts to more positive values, then, at
$i\sim15^{\circ}$ the DOA of the line profiles moves towards zero. At
$i=0^{\circ}$ the DOA of all the phase resolved profiles are zero.  \\

\noindent
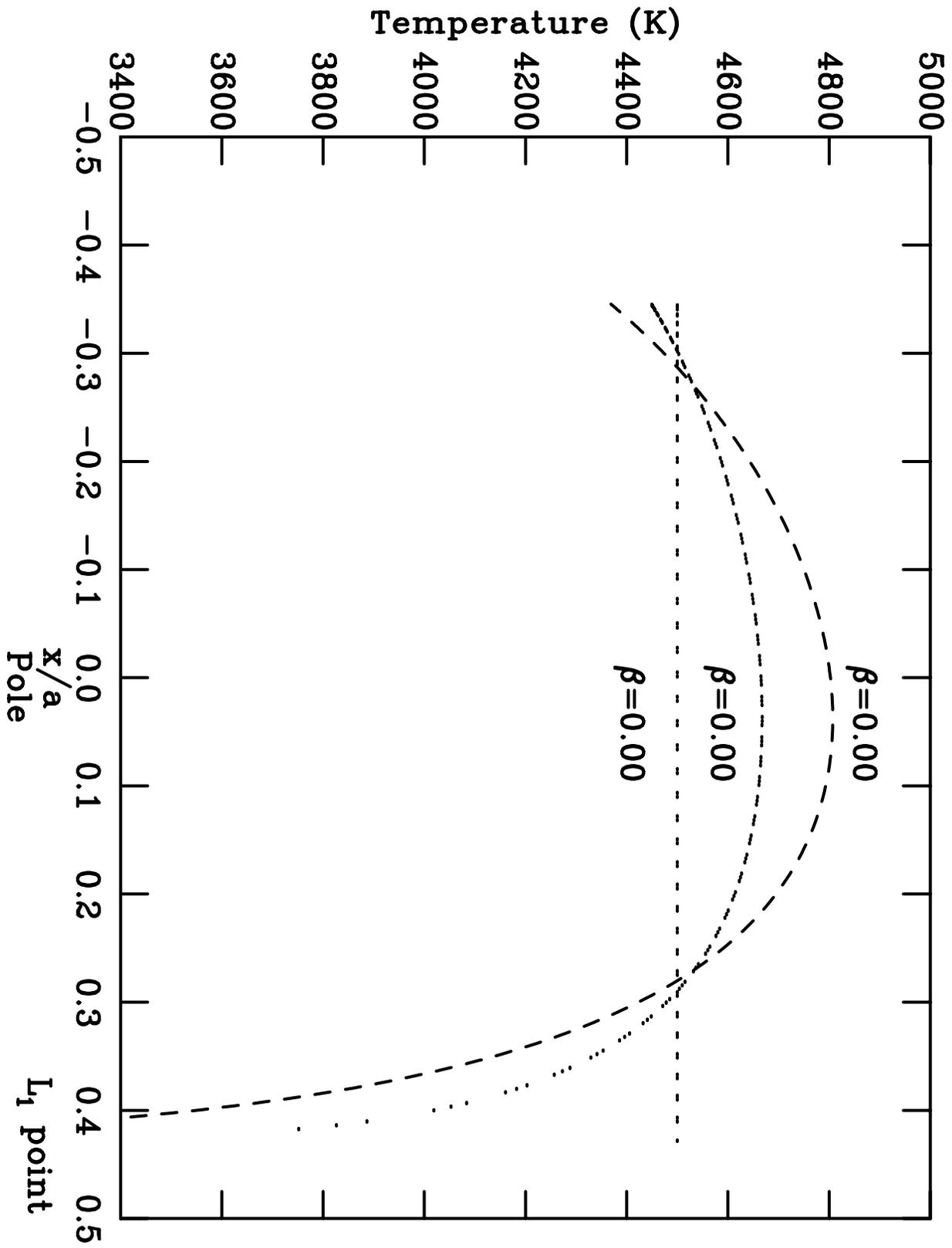
{\bf Figure 4:} The temperature distribution on the surface of the
secondary star along the line of centres of the binary components for
$\beta$=0.0 (dotted line), 0.08 (solid line) and 0.16 (dashed lines). The
distribution has been normalised to a mean temperature of 4500 K. As
$\beta$ increases the temperature of the elements of area near the inner
Lagrangian point ($L_{1}$) also decreases. Thus these elements will emit
less flux compared to regions near the pole of the secondary star.  \\

\noindent
{\bf Figure 5:} The effects of gravity darkening. 5(a) The FWHM and
DOA curves for fixed $i$=20$^{\circ}$. Models for $\beta$=0.00
(stars), 0.08 (circles), and 0.16 (crosses) are shown; 5(b) Same as above
but with fixed $i$=60$^{\circ}$. As one can see the shape of the curves
depends on the value for the binary inclination.  \\

\noindent
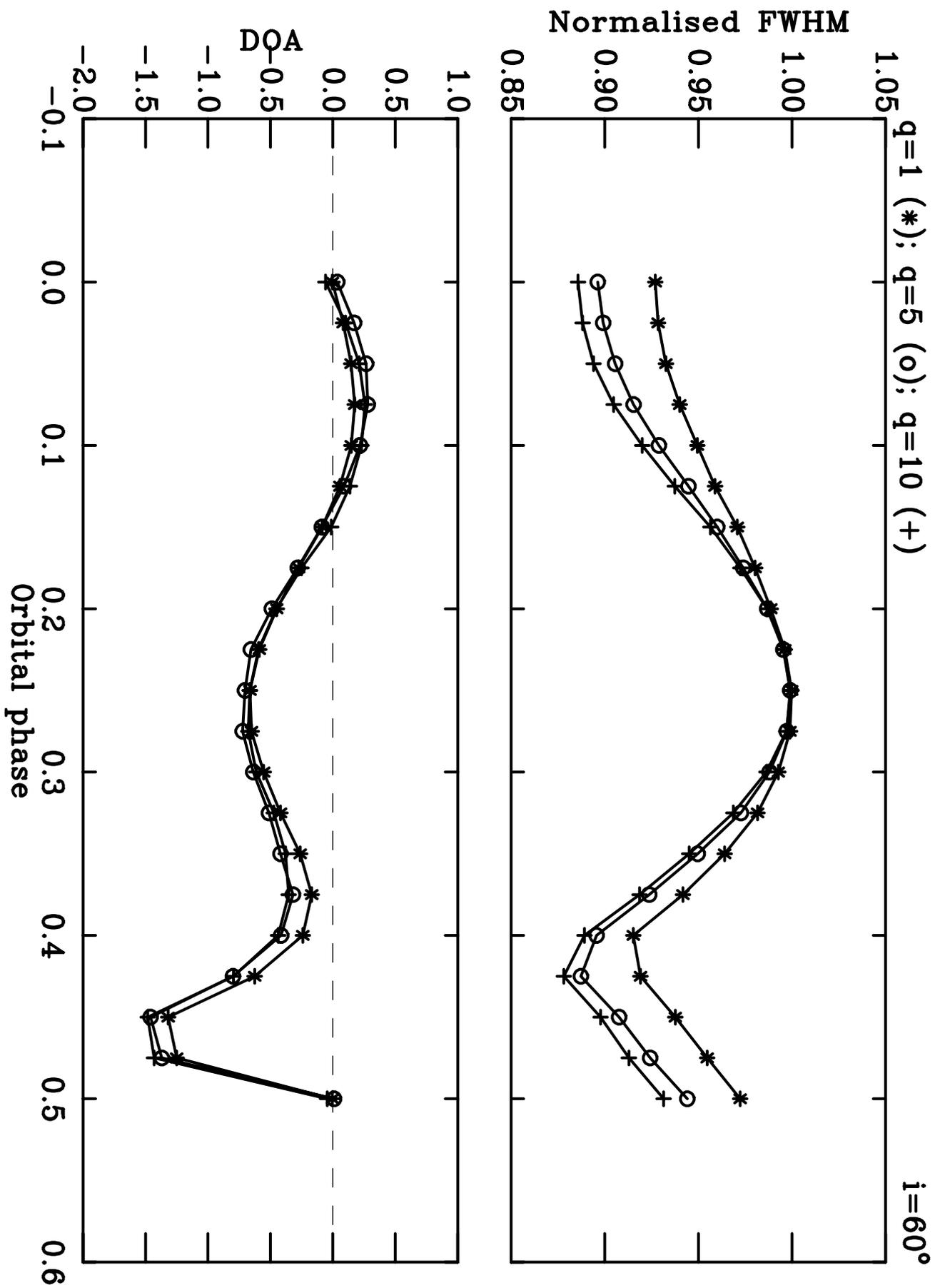
{\bf Figure 6:} The effects of the binary mass ratio on the FWHM and
asymmetry of the line profiles for fixed $i$=60$^{\circ}$. The curves are for
$q=M_{1}/M_{2}$=1 (stars), 5 (circles) and 10 (crosses). The largest
changes occur at small values of $q$. The horizonal dashed line
indicates where the degree of asymmetry of the line is zero.  \\

\noindent
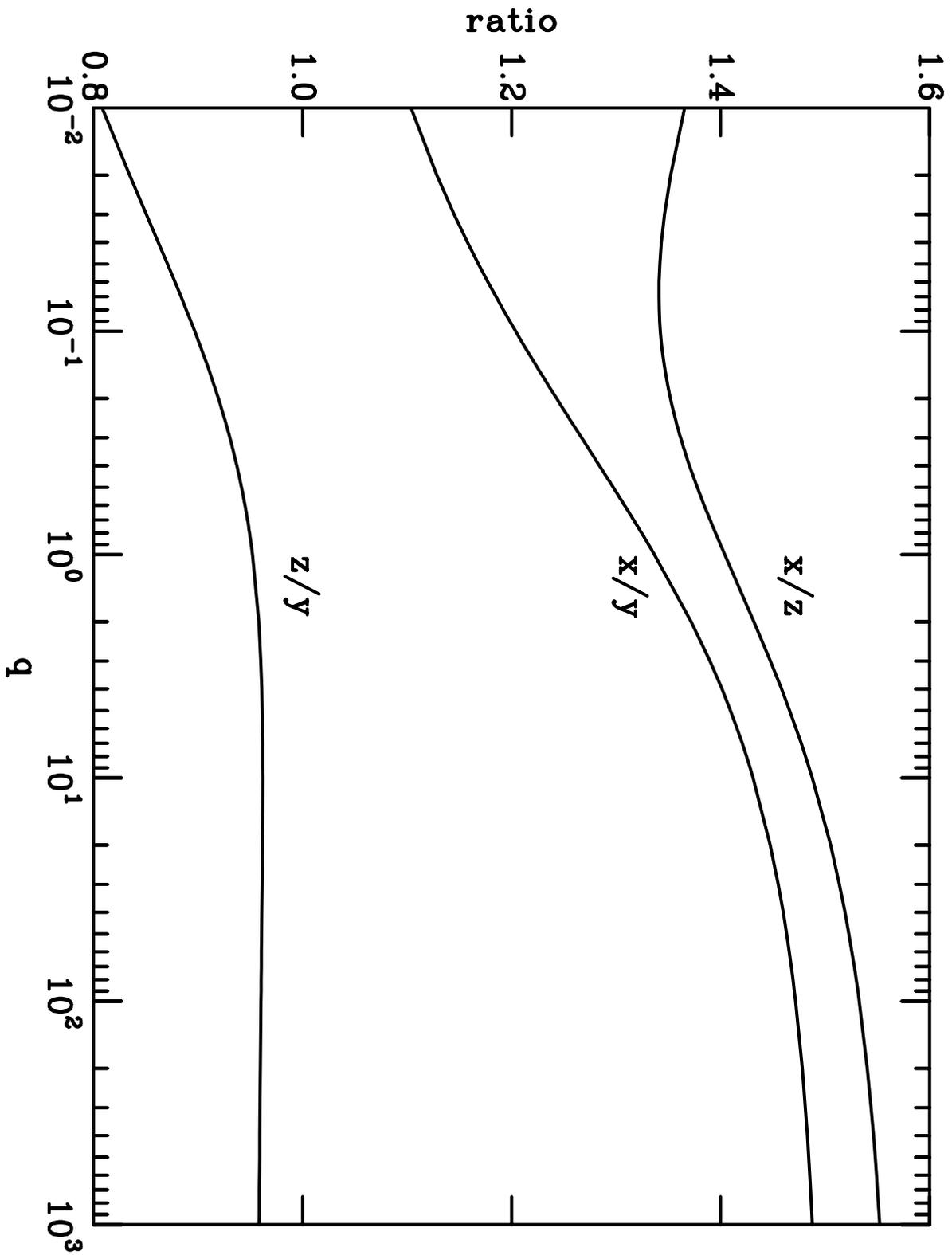
{\bf Figure 7:} The distortion of the secondary star measured by
computing the ratio of the Cartesian coordinates $x/y$, $x/z$ and $z/y$
at the surface of the secondary star $x$ is defined to lie along the
line joining the centre of mass of the two stars, $y$ to lie in the
orbital plane and $z$ perpendicular to the orbital plane.  \\

\noindent
{\bf Figure 8:} The effects of limb darkening. (a) The normalised FWHM and DOA
curves for fixed $i$=30$^{\circ}$. Models for zero ($u$=0.0; crosses) and
full ($u$=1.0; circles) limb darkening are shown; (b) Same as above but
with fixed $i$=50$^{\circ}$.  \\

\noindent
{\bf Figure 9:} (a) Same as Fig 3b but using zonal velocity pattern
across the secondary star. (b) Same as Fig 5b but using a
zonal velocity pattern across the secondary star.
The curves are very similar to those without the velocity pattern 
because the actual velocities are very small (see Martin \& Davey 1995).  \\

\noindent
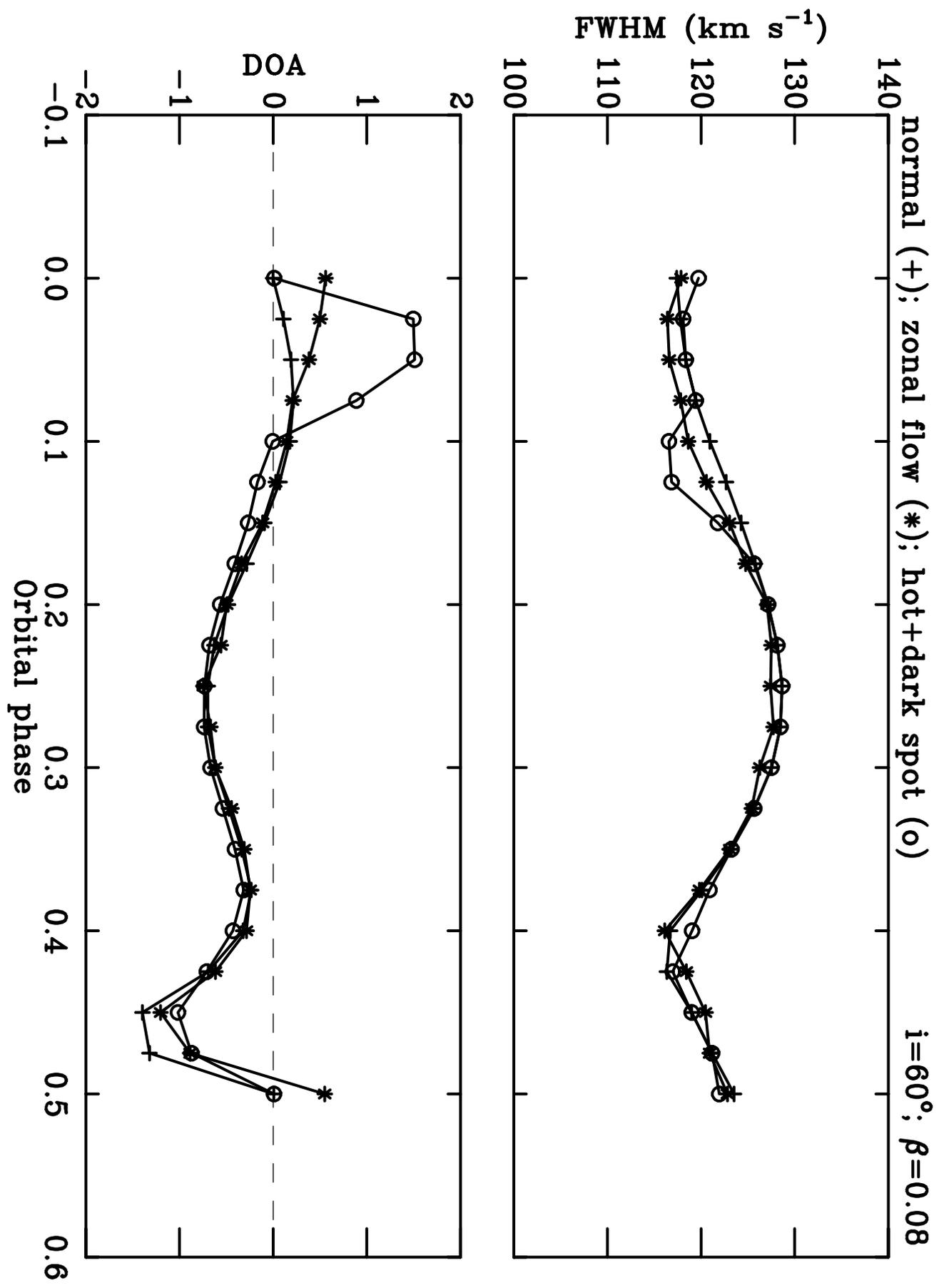
{\bf Figure 10:} The effects of star-spots (dark and hot-spot 10 degrees
in radius; see section 4.2) on the FWHM and DOA curves for a system with
$i$=60$^{\circ}$ and $\beta$=0.08. Also shown for comparison is the case
with (zonal flow) and without (normal) a zonal velocity flow.  \\

\noindent
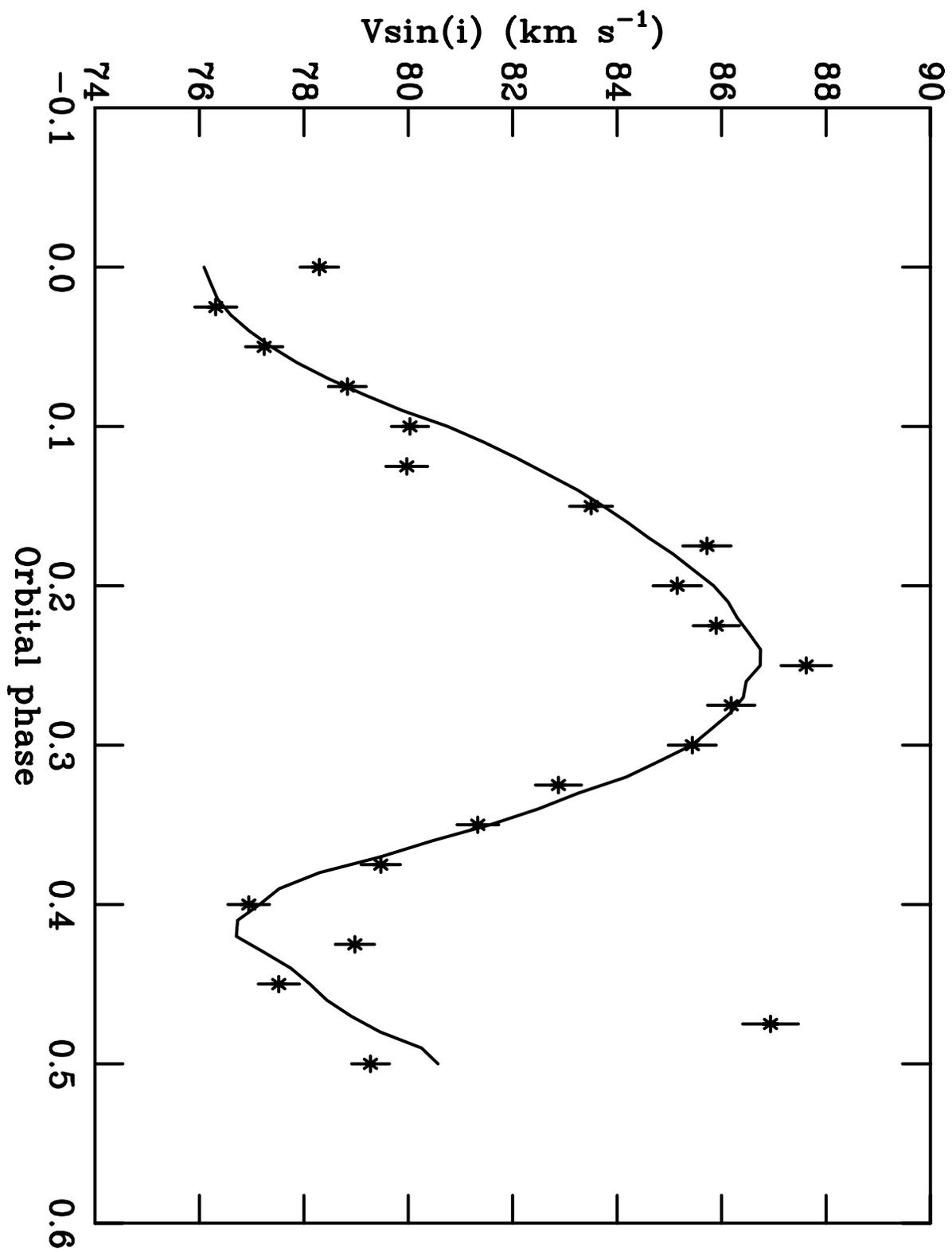
{\bf Figure 11:} The $V_{rot}\sin i$ variations with orbital phase for the
simulated data, obtained by optimally subtracting a template star
broadened using the Gray (1992) profile from the phase resolved data.
Also shown is the best fit obtained using the FWHM variations of the
model line profiles (see section 4.1).  \\

\noindent
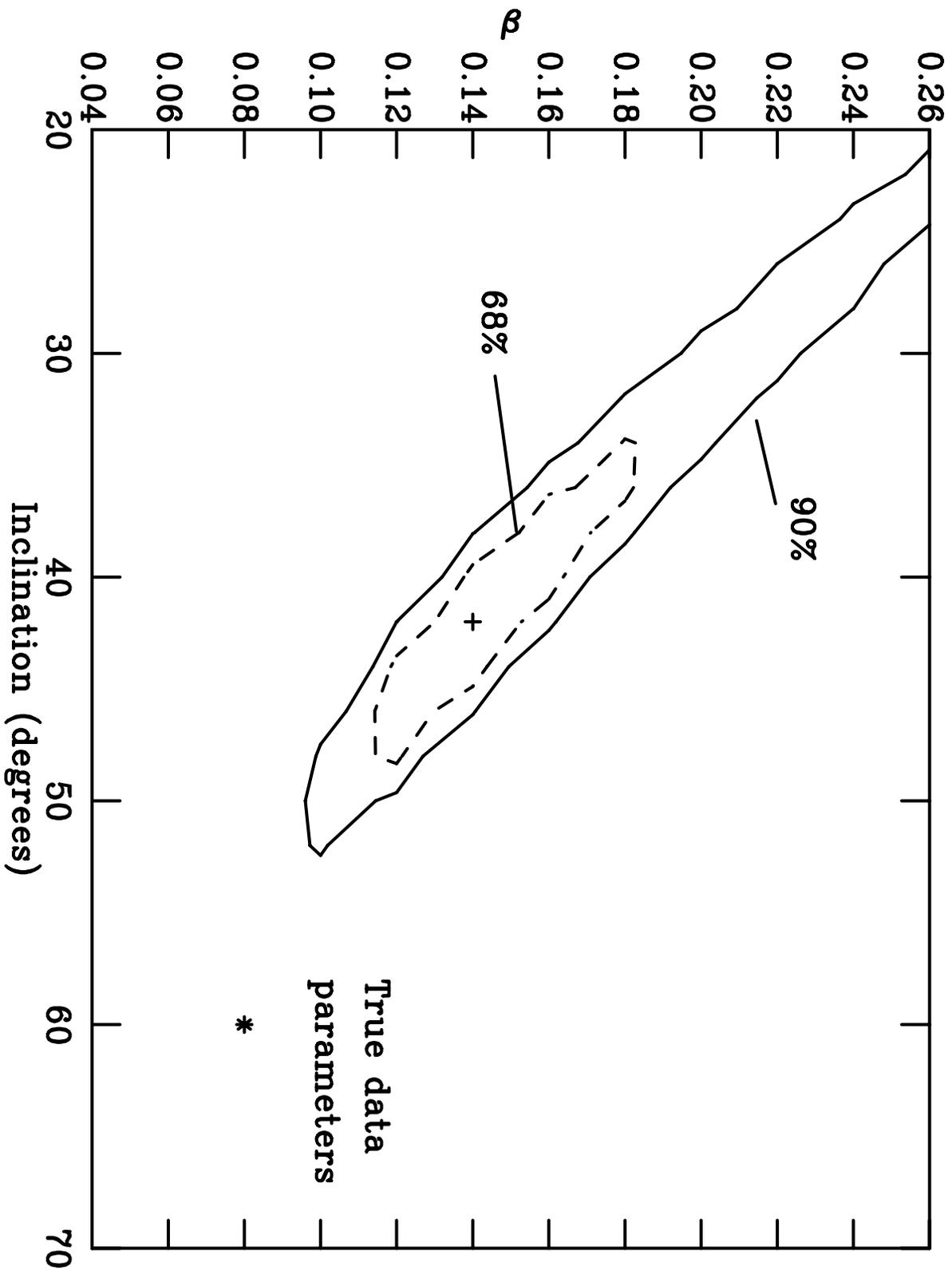
{\bf Figure 12:} Results from fitting the $V_{rot}\sin i$ variations
using the FWHM of the model line profiles. The 68 (dashed line) and 90
(solid line) percent confidence solutions are shown in the $\beta-i$
plane. The best model fit is marked at $i$=42$^{\circ}$ and $\beta$=0.14
(cross) and the true solution of the simulated data is also marked at
$i$=60$^{\circ}$ and $\beta$=0.08 (star).  \\

\noindent
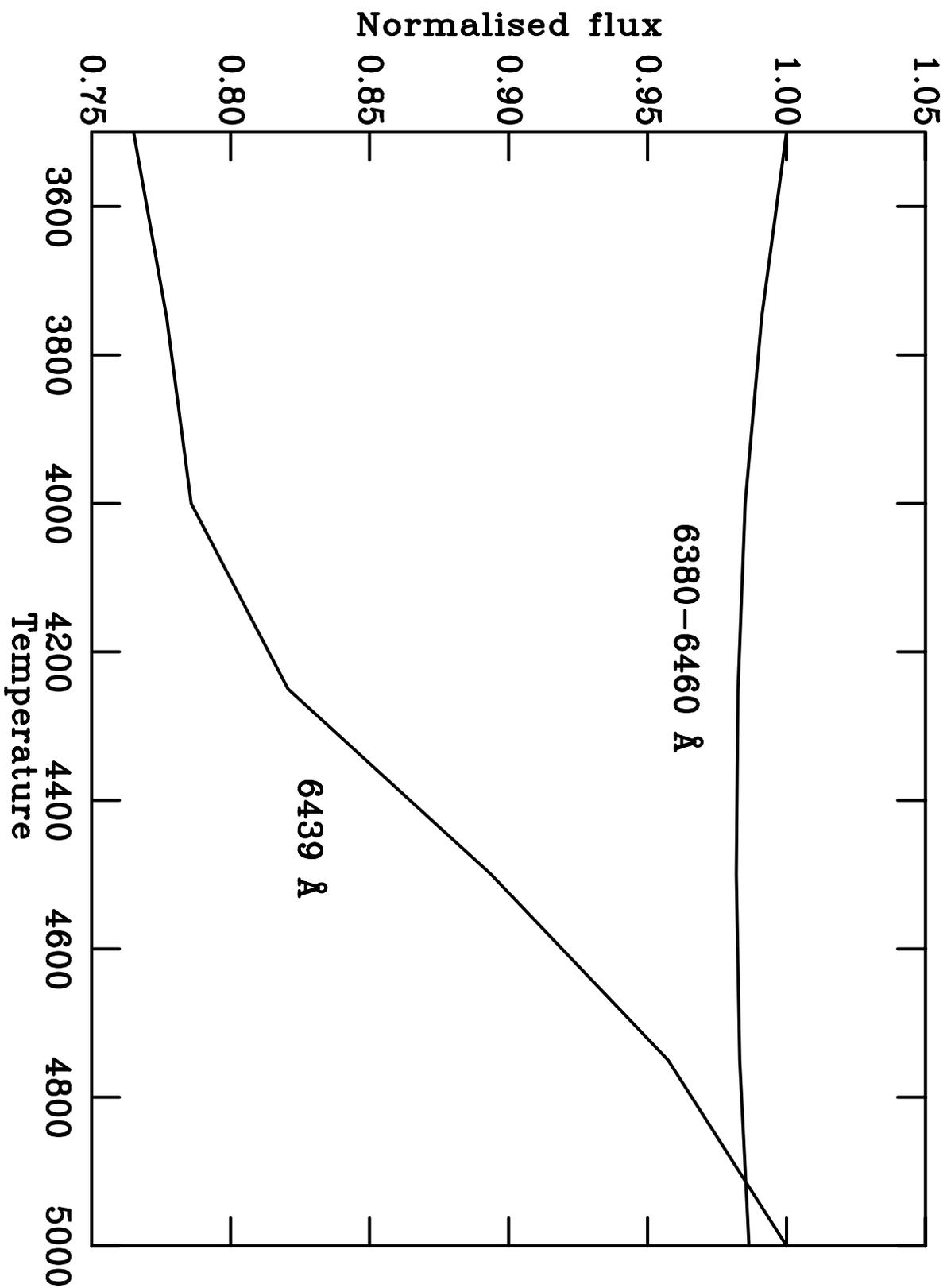
{\bf Figure 13:} The normalised specific intensity as a function of
temperature (log~$g$=4.5) in the Ca $\sc i$ 6439 \AA\ absorption line
compared to the total specific intensity in the spectral range 6380--6460
\AA. As one can see the single line has a much stronger dependence on 
temperature and so is more sensitive to the fitting procedure outlined in
section 5.2.  \\

\noindent
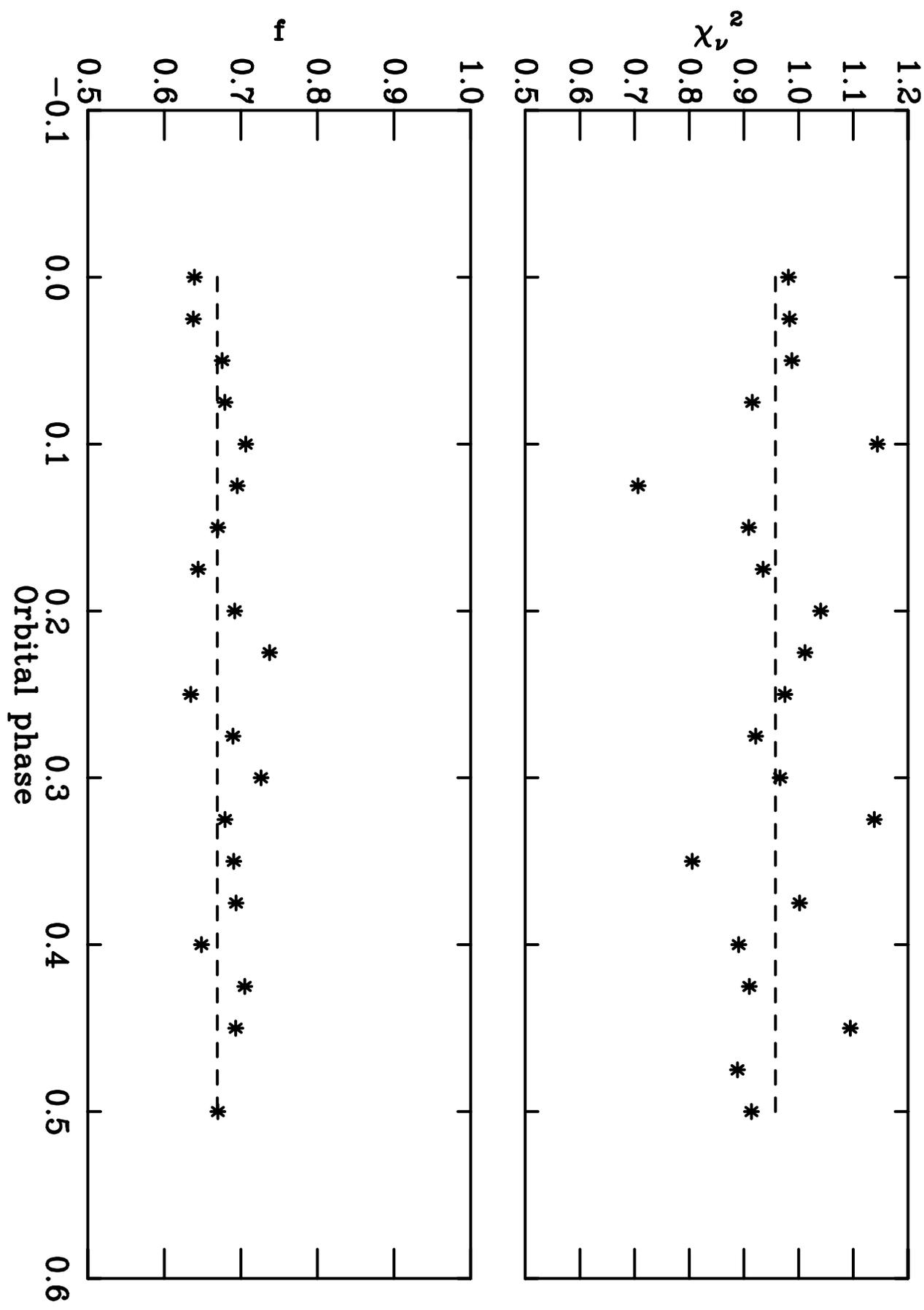
{\bf Figure 14:} Results for fitting the Ca $\sc i $ line profile with
the model described in section 3. The 68 percent confidence solution is
shown in the $\beta-i$ plane. The best model fit is marked
$i$=66$^{\circ}$ and $\beta$=0.06 (cross) and the true solution is also
marked at $i$=60$^{\circ}$ and $\beta$=0.08 (star). \\

\noindent
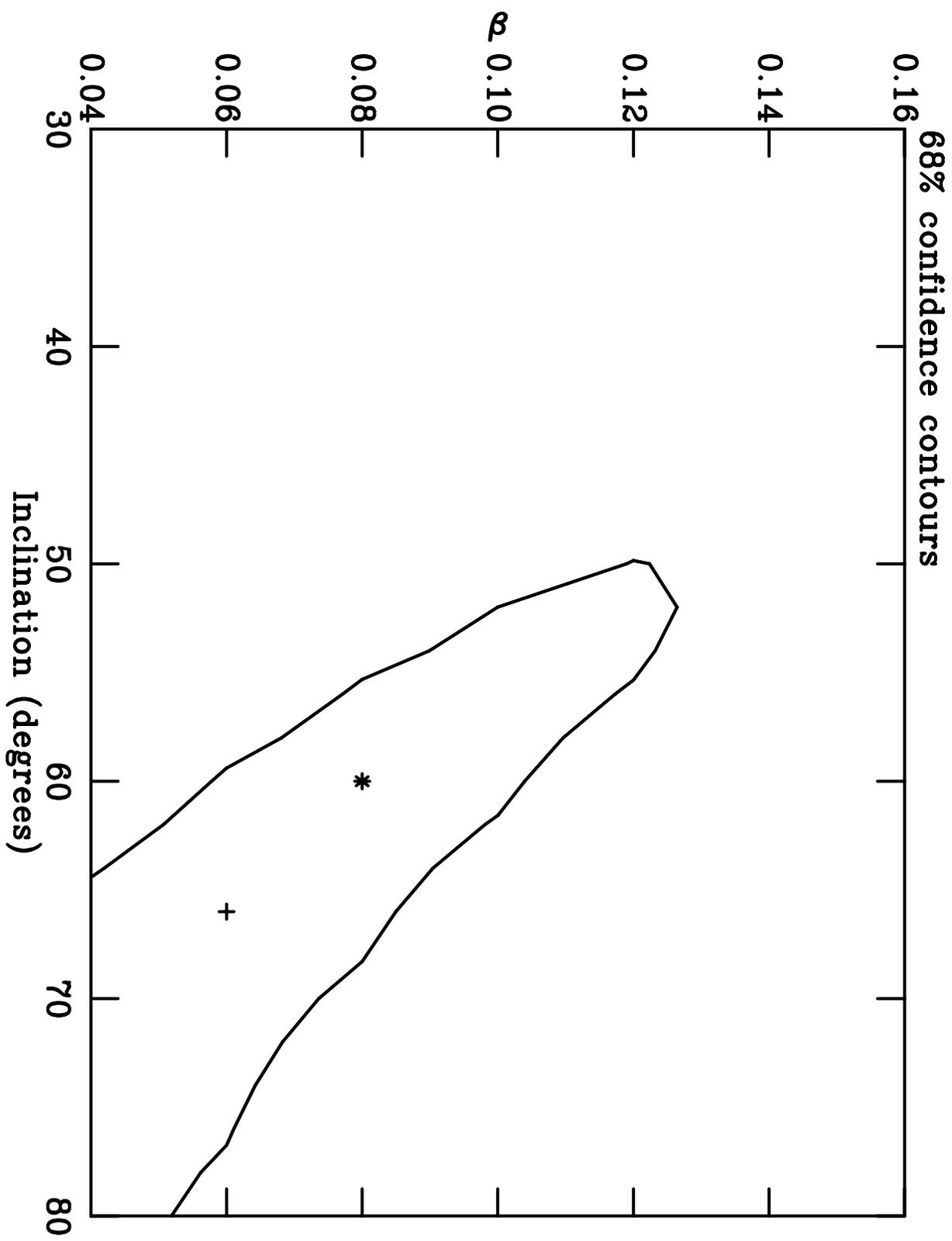
{\bf Figure 15:} The results for the optimal subtraction of the simulated
data and the best fit model. The top panel shows the $\chi^{2}_{\nu}$ for
each phase resolved spectrum. The bottom panel shows the fraction of
light arising from the secondary star determined by the phase resolved
optimal subtraction. \\

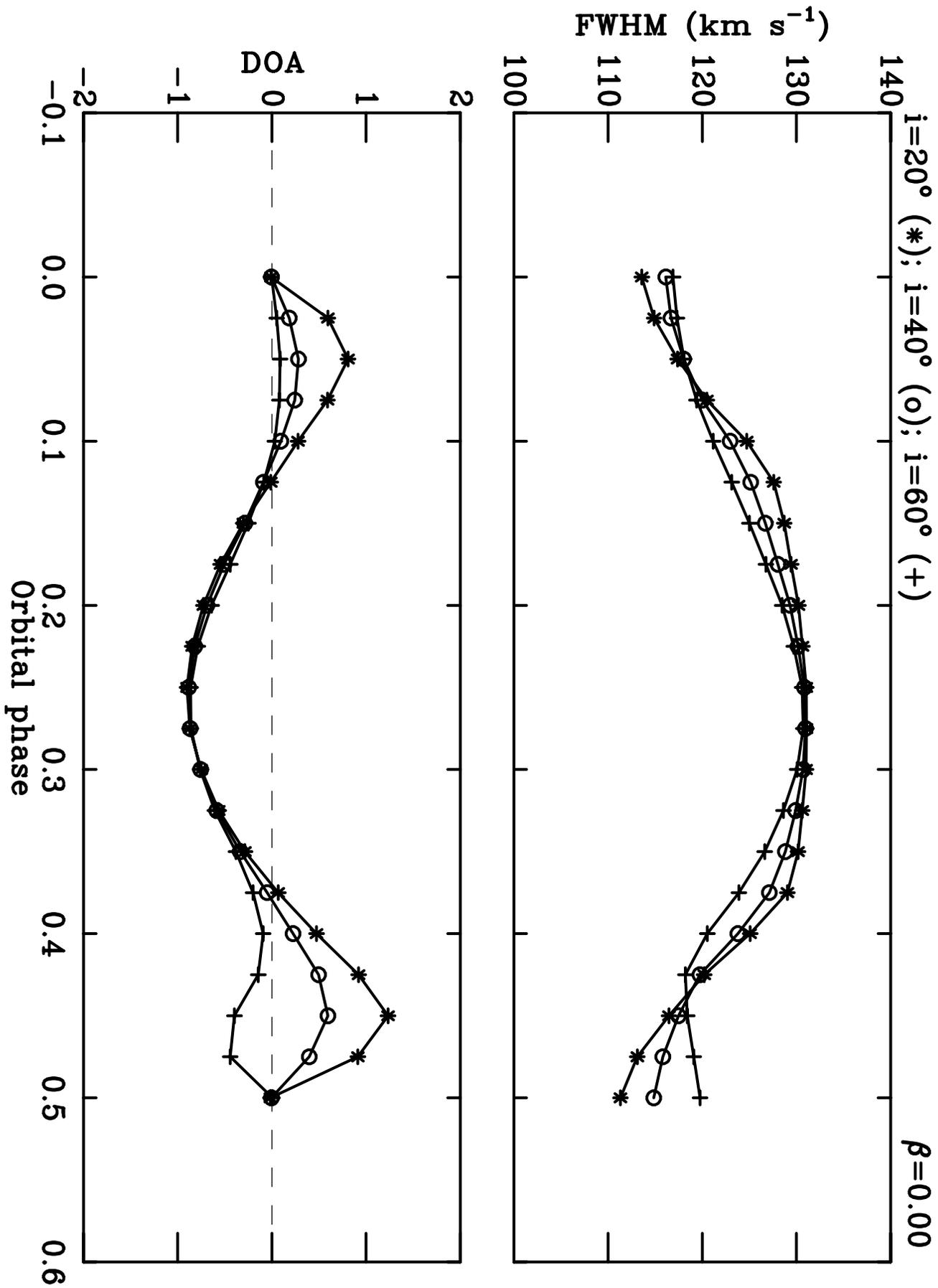
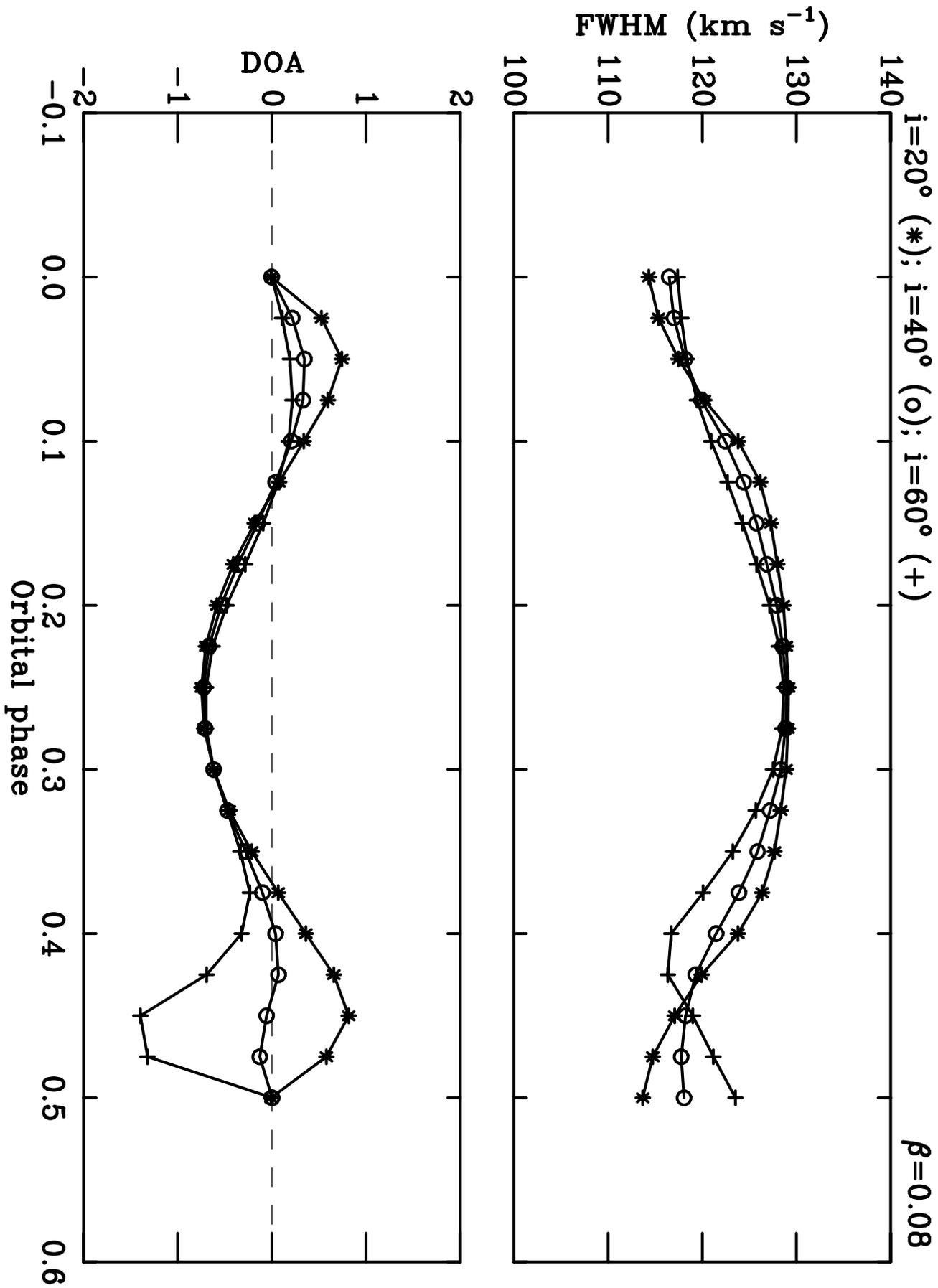
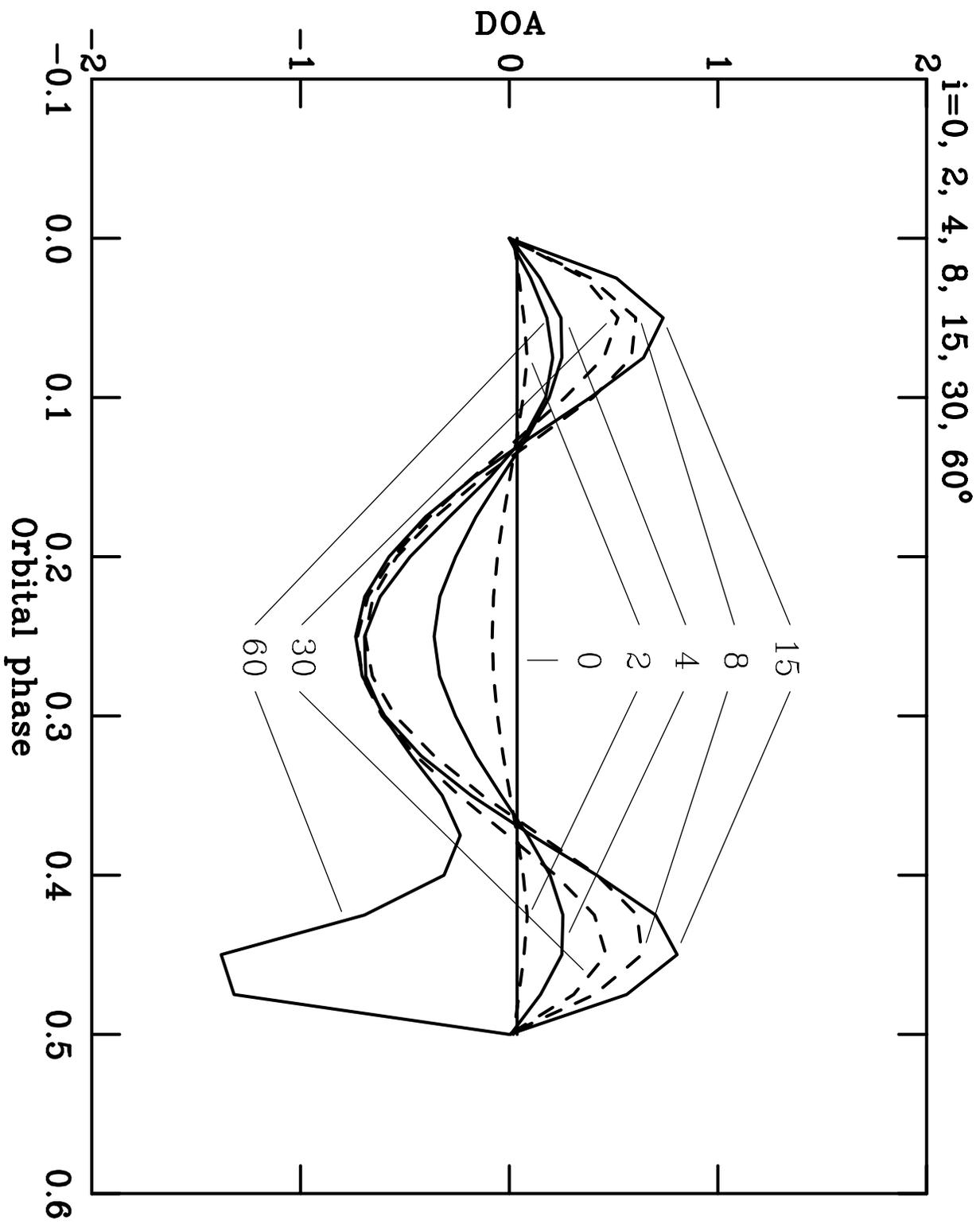
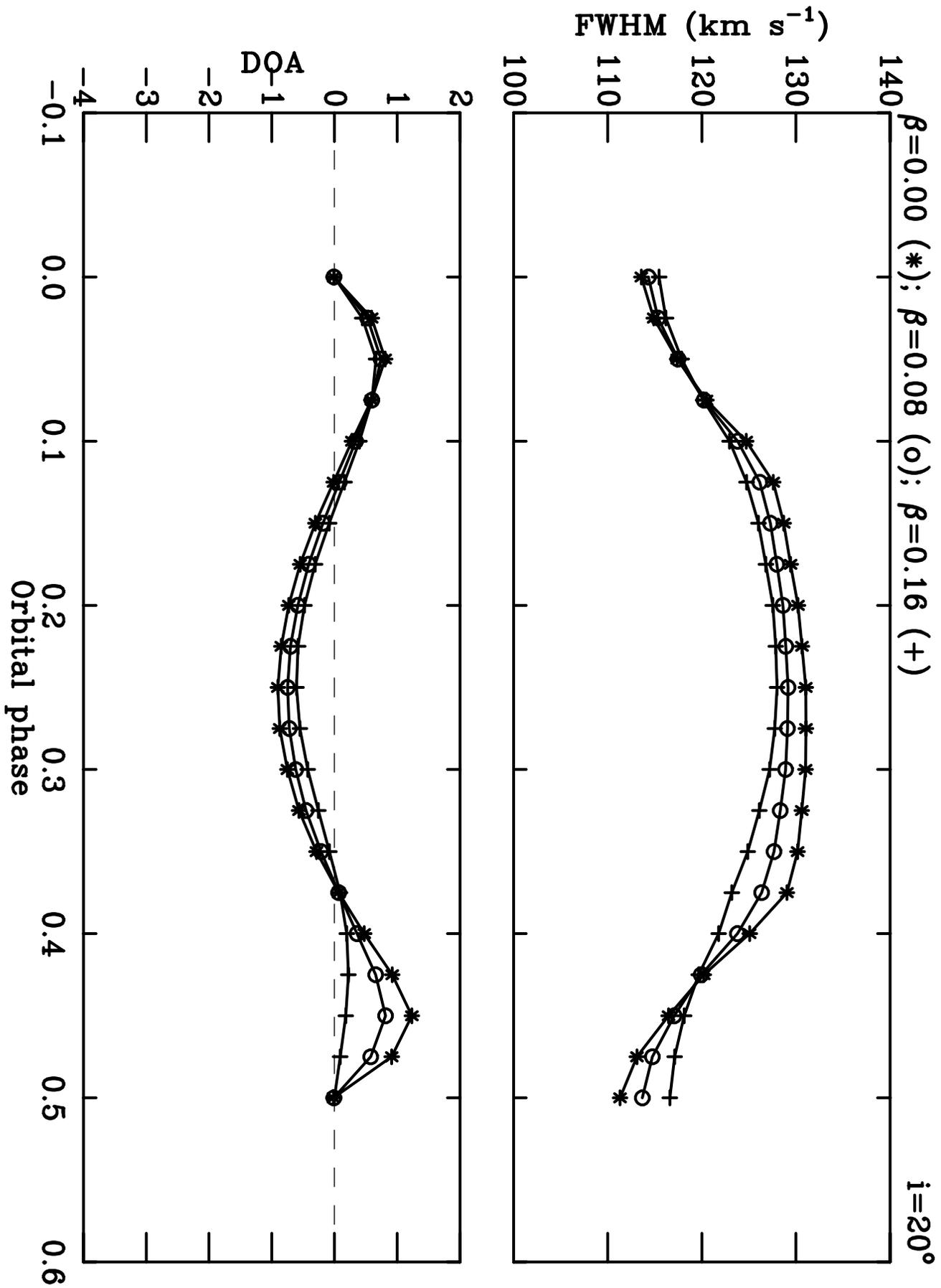
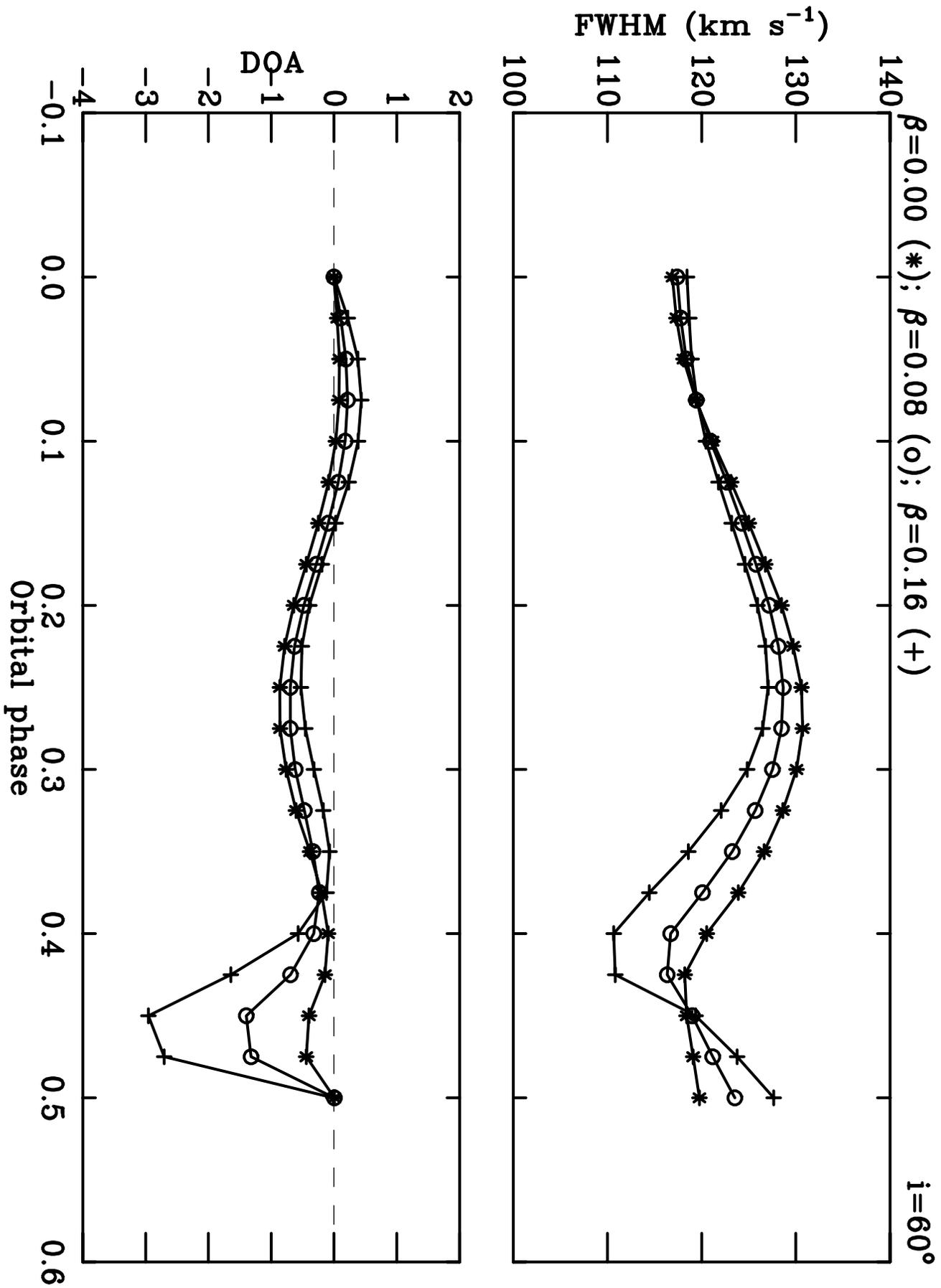
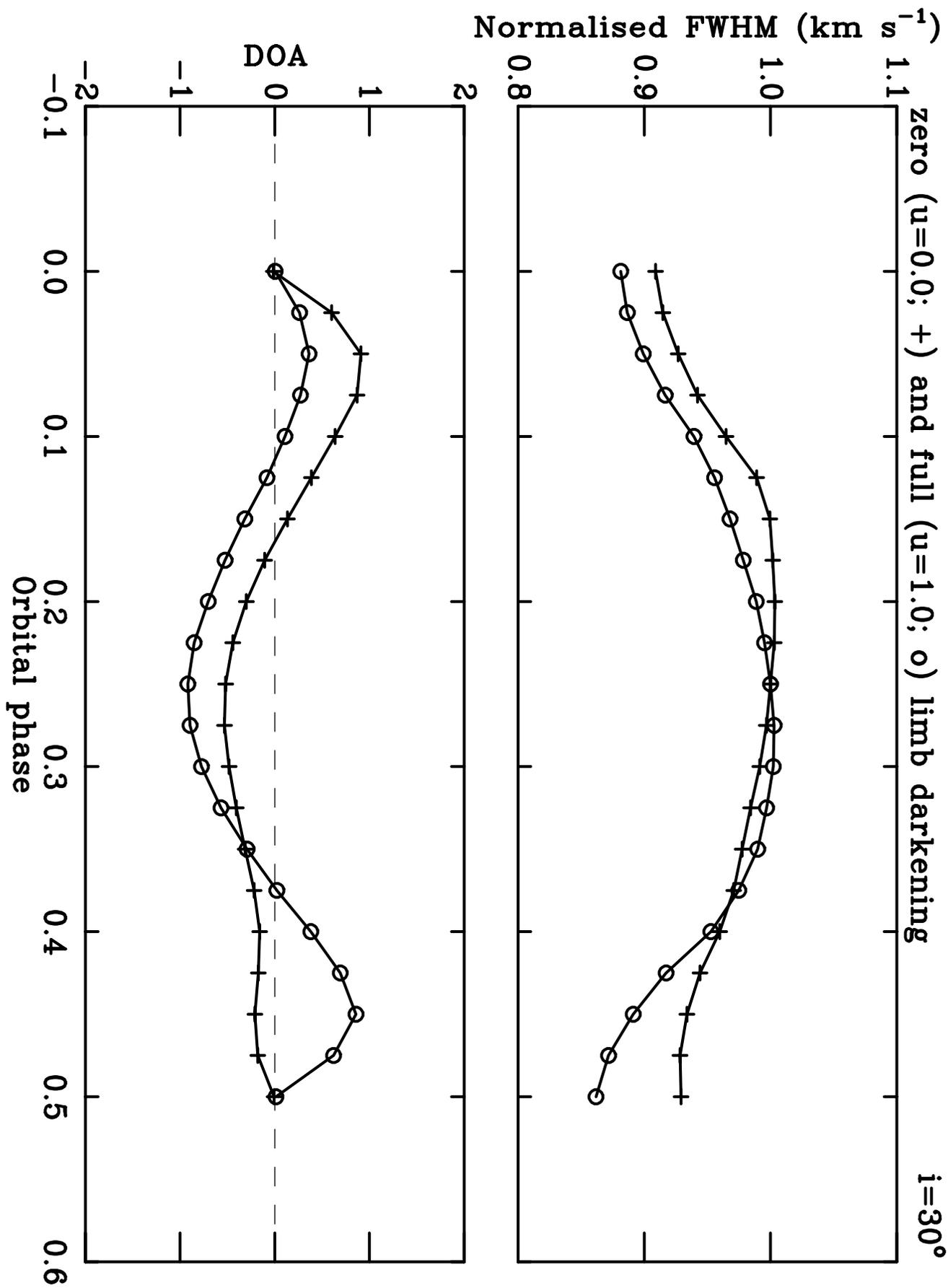
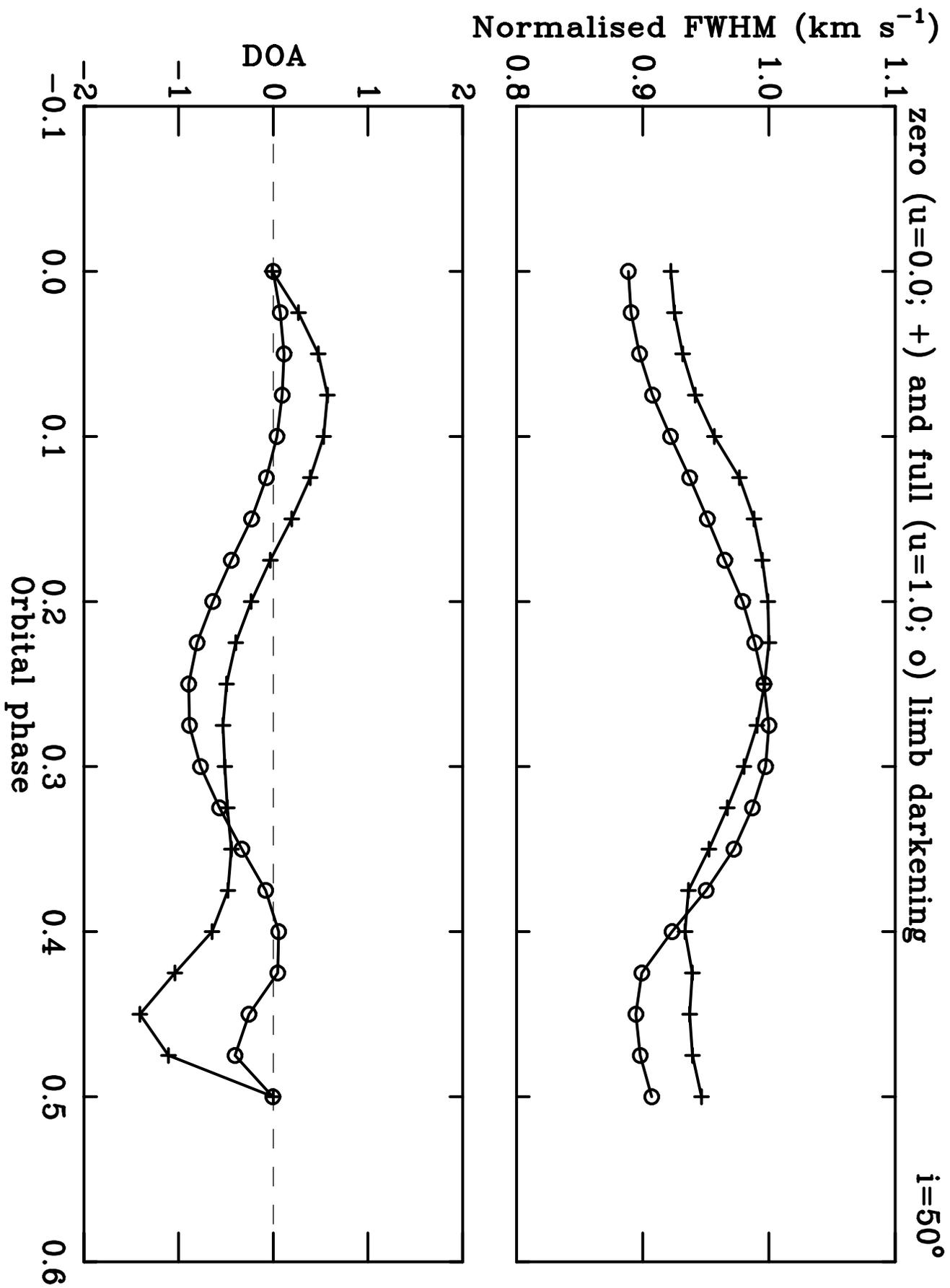
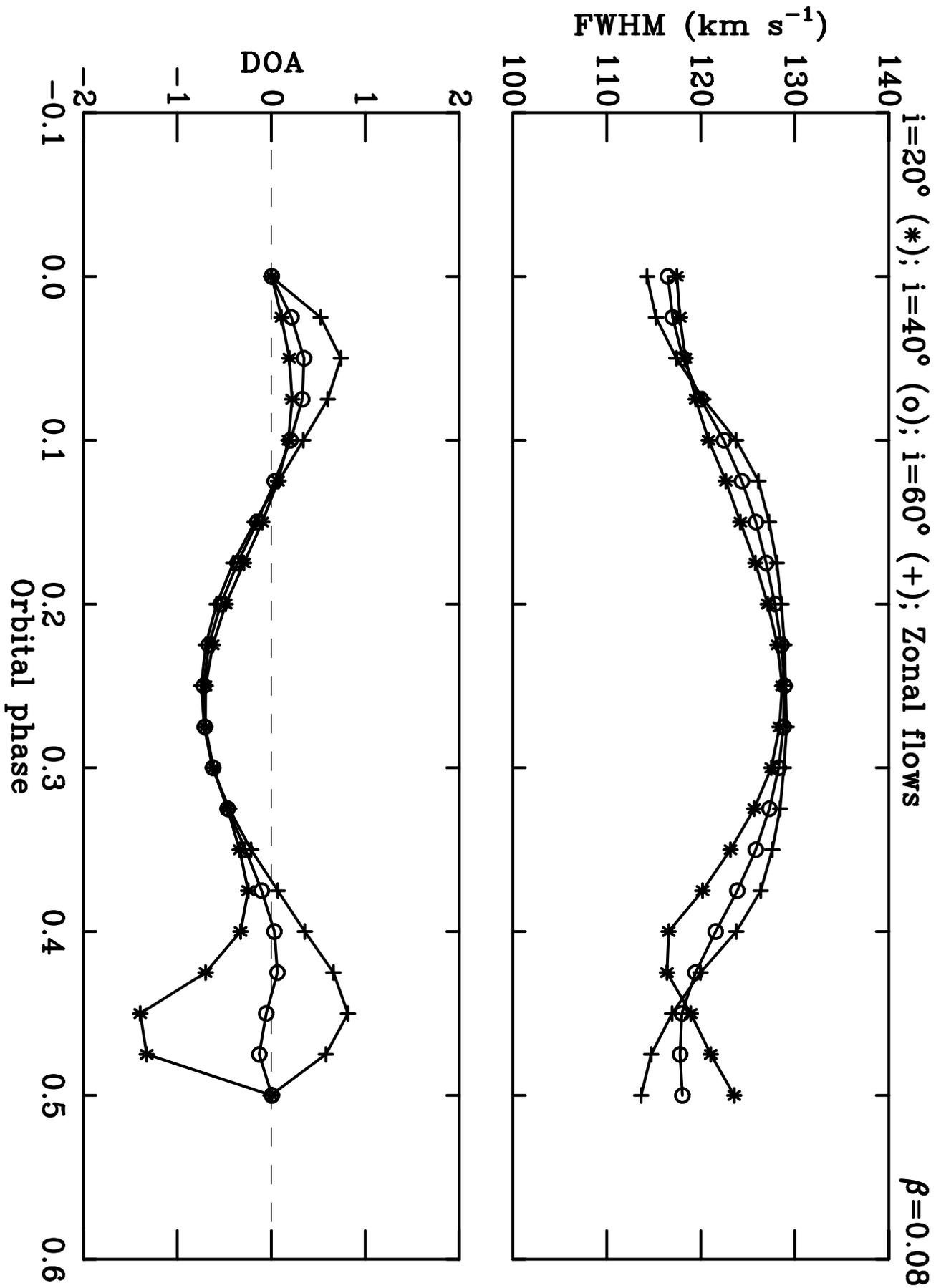
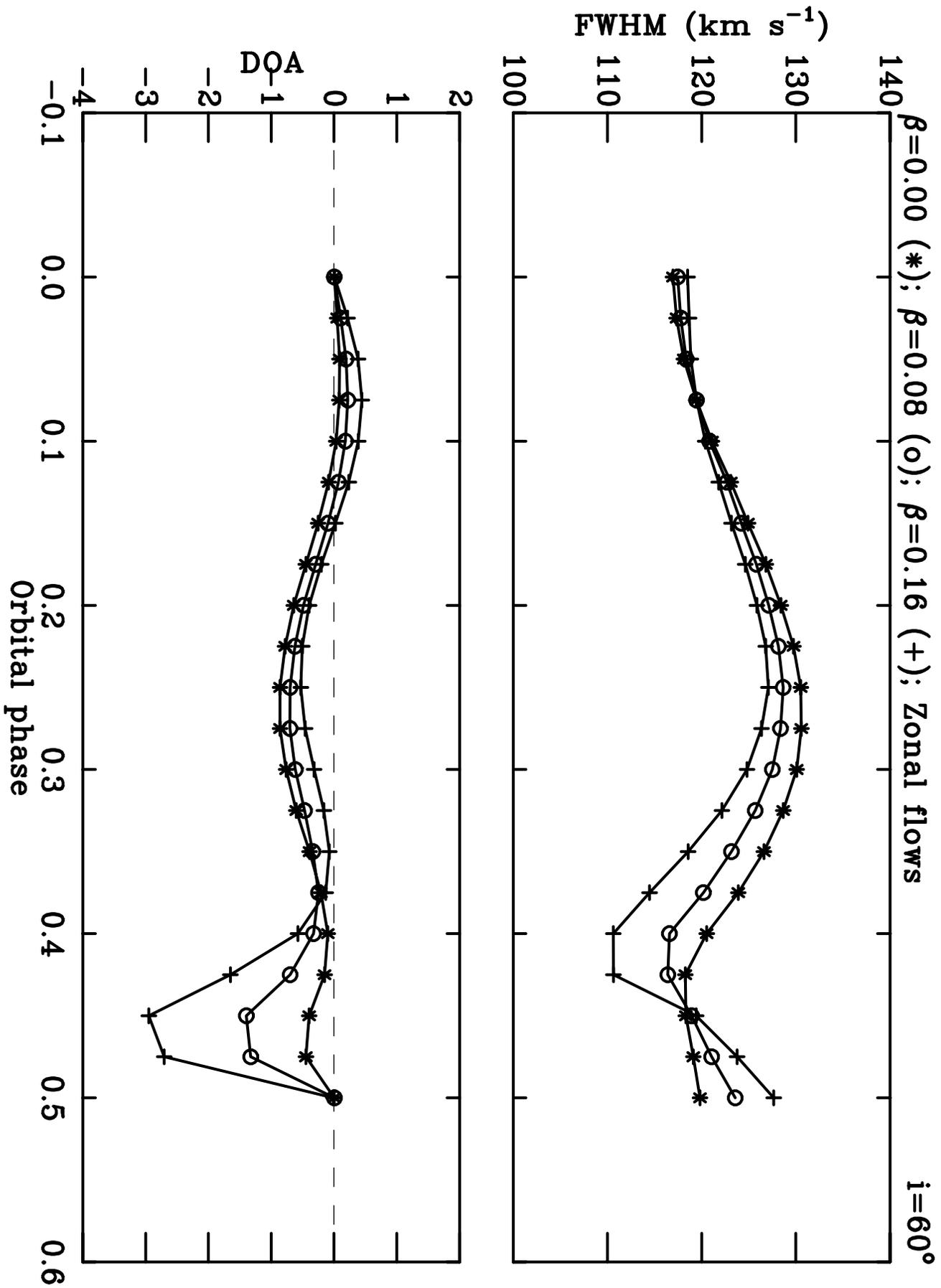
\end{document}